\documentclass[twocolumn,aps,prl,superscriptaddress,floatfix]{revtex4-2}

\usepackage{amsmath}
\usepackage{graphicx}
\usepackage{physics}
\usepackage{bbold}
\usepackage[colorlinks=true, allcolors=blue]{hyperref}

\begin{document}

\title{Theory of potential impurity scattering in pressurized superconducting La$_3$Ni$_2$O$_7$}
\author{Steffen B\"otzel}
\affiliation{Theoretische Physik III, Fakult\"at für Physik und Astronomie, Ruhr-Universit\"at Bochum,
  D-44780 Bochum, Germany}
  \author{Frank Lechermann}
  \affiliation{Theoretische Physik III,  Fakult\"at für Physik und Astronomie, Ruhr-Universit\"at Bochum,
  D-44780 Bochum, Germany}

\author{Takasada Shibauchi}
\affiliation{Department of Advanced Materials Science, The University of Tokyo, Kashiwa, Chiba 277-8561, Japan}

\author{Ilya M. Eremin}
\affiliation{Theoretische Physik III,  Fakult\"at für Physik und Astronomie, Ruhr-Universit\"at Bochum,
  D-44780 Bochum, Germany}

\begin{abstract}
Recently discovered high-T$_c$ superconductivity in pressurized bilayer nickelate La$_3$Ni$_2$O$_7$ (La-327) is believed to be driven by the non-phononic repulsive interaction. Depending on the strength of the interlayer repulsion, the symmetry of the superconducting order parameter is expected to be either $d$-wave or sign-changing bonding-antibonding $s_{\pm}$-wave. Unfortunately, due to the need of high pressure to reach superconducting phase, conventional spectroscopic probes to validate the symmetry of the order parameter are hard to use. Here, we study the effect of the point-like non-magnetic impurities on the superconducting state of La-327 and show that $s_{\pm}$-wave and $d$-wave symmetries show a very different behavior as a function of impurity concentration, which can be studied experimentally by irradiating the La-327 samples by electrons prior applying the pressure. While $d-$wave superconducting state will be conventionally suppressed, the $s_{\pm}$-wave state shows  more subtle behavior, depending on the asymmetry between bonding and antibonding subspaces. For the electronic structure, predicted to realize in La-327, the $s_{\pm}-$wave state will be robust against complete suppression and the transition temperature, $T_c$ demonstrates a transition from convex to concave behavior, indicating a crossover from $s_{\pm}$-wave to $s_{++}$-wave symmetry as a function of impurity concentration. We further analyze the sensitivity of the obtained results with respect to the potential electronic structure modification.        
\end{abstract}

\maketitle

\section{Introduction}

The discovery of high-temperature superconductivity in pressurized La$_3$Ni$_2$O$_7$ (La-327)  ~\cite{sun23,JunHou:117302,zhang2023high,zhou2023evidence,zhang2023effects,wang2023structure,wang2023pressure,dong2023visualization} and La$_2$PrNi$_2$O$_7$ \cite{WangNature24}, is remarkable not only because of the observed high superconducting transition temperature of about $T_c \sim 80$~K but also due to the peculiar electronic structure of this bilayer Ruddelsden-Popper (RP) perovskite, which is different from that of hole-doped thin films of superconducting infinite-layer and reduced multilayer nickelates~\cite{Li2019,Pan2021,Osada2020}. Considering La-327 as RP bilayer systems yields a formal Ni $3d^{7.5}$ (or $3d^8$ when considering ligand-hole physics~\cite{lechermann2023electronic}) electronic configuration with both Ni-$e_{g}$ orbitals crossing the Fermi level. The low-energy physics in this system is governed by the multiorbital and bilayer effects with strong hybridization between the Ni-$d_{z^2}$ and the apical O-$p_z$ orbitals \cite{zhang2023electronic}. The multiorbital structure seems to be also one of the key differences between La-327 and the bilayer cuprate superconductors where Cu$^{2+}$ ions with $3d^9$ configuration possess only one valence hole in the $3d_{x^2-y^2}$ orbital, whereas the Ni ion has unpaired valence electrons (holes) in both the $3d_{x^2-y^2}$ and $3d_{z^2}$ orbitals.
Various Hubbard-Hund-type or $t-J$-like models have already been proposed to capture the superconducting and normal state properties of this multiorbital system~\cite{lechermann2023electronic,liu2023role,qin2023high,huang2023impurity,oh2023type,qu2023bilayer,ryee2023critical,tian2023correlation,liao2023electron,kaneko2023pair,LuoModel,chen2023orbital,jiang2023high,shen2023effective,yang2023minimal,wu2023charge,luo2024high}. 

Within the variety of model considerations, one of the most interesting theoretical questions concerns the interplay between the intralayer and the interlayer Cooper pairing \cite{dagotto1992superconductivity}, which yields a competition between the $s_{\pm}$-wave symmetry of the superconducting order parameter, driven mostly by the interlayer interaction~\cite{liu2023role,lechermann2023electronic,qin2023high,zhang2023trends,yang2023possible,luo2024high,oh2023type,qu2023bilayer,zhang2023structural,huang2023impurity,heier2023competing,yang2023strong,sakakibara2024possible} and the $d_{x^2-y^2}$-wave or the $d_{xy}$-wave symmetries of the superconducting order parameters, driven mostly by the intralayer interaction, respectively~\cite{lechermann2023electronic,liu2023role,jiang2023high,heier2023competing,lu2023interlayer}. 

The bilayer structure of the La-327 allows to separate the electronic states into bonding and antibonding combination with respect to the layer index. The strong interlayer hybridization (hopping) mediates particularly strong splitting of the bonding and antibonding $3d_{z^2}$ orbitals. While the latter is above the Fermi level, the former forms a flat band (frequently denoted as $\gamma$ band). Angle-resolved photoemission spectroscopy (ARPES) at low temperatures sees that band slightly below the Fermi level at ambient pressure \cite{yang2024orbital,li2024electronic}. It has been argued that the $\gamma$-band crosses the Fermi energy at the high-pressure phase, as supported by a clear drop of the Hall coefficient indicating an increase of hole carrier density~\cite{zhang2024high,zhouguo23}. The hybridization of the $d_{x^2-y^2}$-orbitals between the two layers is somewhat smaller due to their in-plane character and also vanishes along the diagonal of the Brillouin-Zone making bonding-antibonding $d_{x^2-y^2}$-mostly bands degenerate along this direction~\cite{liu2023role}.

Due to the flatness of the $\gamma$-band, many theory works pointed out its potential importance for superconductivity irrespective of the type of the dominant superconducting instability~\cite{liu2023role,lechermann2023electronic,qin2023high,zhang2023trends,yang2023possible,luo2024high,oh2023type,qu2023bilayer,zhang2023structural,huang2023impurity,heier2023competing,yang2023strong,sakakibara2024possible}. 
Whether the system chooses $s_{\pm}$-wave or $d$-wave symmetry of the superconducting order parameter depends more on the relative strength of the interlayer versus intralayer antiferromagnetic spin fluctuations~\cite{Boetzel2024} yet the presence of the $\gamma$-band may affect this competition. 

The current experimental challenge is to have a reliable experimental probe, which would allow to distinguish between the $d$-wave and unconventional $s_{\pm}$-wave symmetries in bilayer nickelates. Unfortunately, conventional spectroscopic techniques are hard to use under high pressures although they could provide reliable hints if one for example studies impurity-induced bound states~\cite{HuangPRB23}, frequency-dependent spin susceptibility~\cite{Boetzel2024}, or Andreev reflections~\cite{yang2024possible}. At the same time, multiband $d$-wave and unconventional $s_{\pm}$-wave symmetries are expected to react differently to the non-magnetic point-like impurity scattering, which was investigated previously in detail in iron-based superconductors~\cite{Efremov2011,Hirschfeld2015,Korshunov2016}. The latter can be achieved by irradiating high-energy light particles (such as electrons, neutrons, or protons) to the samples \cite{Mizukami2014}, before the application of pressure. 

One should note, however, that the sublattice character of the electronic states introduces further caveat as it restricts the possible scatterings due to impurities, making them sometimes only weakly pair-breaking for spin-singlet superconducting phases. For example, it was recently shown to be the case for the kagome lattice \cite{HolbaekPRB23}. For the bilayer system, one has to separately average over impurities in each layer of a bilayer sandwich, which results into an averaging over bonding and antibonding subspaces as was first noted in Ref.~\cite{Hofmann1990} for a constant and positive gap functions for bonding and antibonding bands. This naturally raises the question on the role of sublattice degrees of freedom in bilayer systems such as La-327 with potential interlayer $s_\pm$-wave.

In this manuscript, we analyze the $T_c$ suppression in superconducting La-327 due to point-like non-magnetic impurities focusing on the proposed interlayer constant $s_\pm$-wave Cooper-pairing versus a $d_{x^2-y^2}$ (or $d_{xy}$)-wave Cooper pairings scenarios. We first consider a simple cuprate-like bilayer model with one bonding and one antibonding band of 3$d_{x^2-y^2}$-orbital character and derive compact analytic expressions for the bonding and antibonding superconducting gap functions, renormalized by impurity scattering. For the half-filled case both $s_{\pm}$- and $d$-wave superconducting states are equally suppressed due to impurities following the Abrikosov-Gor'kov (AG) pair-breaking behavior \cite{abrikosov1961zh}. The deviation from half-filling reduces $T_c$ suppression for the $s_{\pm}$-wave case allowing it to distinguish from the $d$-wave case. In particular, the $T_c$ suppression is of AG-behavior if bonding and antibonding band similarly contribute at the Fermi surface and decreased if one or the other dominates at the Fermi surface. This decrease happens because impurities first induce an intralayer $s$-wave component and upon further increase of the impurity concentration enforce a $s_\pm \rightarrow s_{++}$ transition of the gap structure, changing the $T_c/T_{c0}$ curve from a concave to a convex shape. Therefore, well away from half-filling $T_c$ decreases but may remain finite for the bonding-antibonding $s$-wave case and the saturation of the $T_c$ indicates a transition from the $s_{\pm}$-wave to $s_{++}$-wave superconductivity. We then analyze the $T_c$ behavior in the bilayer model of La-327 and show that $d$-wave and $s_{\pm}$-wave symmetry scenarios can be clearly separated by studying the evolution of 
$T_c$ as a function of point-like disorder concentrations. Furthermore, for the $s_{\pm}$-wave case we argue that the electronic structure modification of the normal state also affects the suppression rate and the crossover concentration at which $s_{\pm}$-wave state transforms into $s_{++}$-wave superconducting state. For example, the presence of the $\gamma$-band near the Fermi level enhances the asymmetry of the bonding-antibonding subspaces of the low-energy electronic structure and makes the $s_{\pm}$-wave state more robust against adding non-magnetic impurities.

\section{Disordered bilayer model}

We consider a bilayer Hamiltonian with point-like non-magnetic impurities $\mathcal{H} = \mathcal{H}_0 + \mathcal{H}_{\text{int}} + \mathcal{H}_{\text{dis}}$, where the interaction term gives rise to an instability towards unconventional $s_\pm$- or $d$-wave superconductivity, respectively. The non-interacting part is given by
\begin{equation}
   \mathcal{H}_{0} = \sum_{\mathbf{k}}\sum_{l_1,l_2}\sum_{o_1,o_2} (\hat{H}_0)_{l_1o_1;l_2o_2}(\mathbf{k}) c^{\dagger}_{l_1o_1}(\mathbf{k})c_{l_2o_2}(\mathbf{k}),  
\end{equation}
where $c^{\dagger}_{l,o}(\mathbf{k})$ creates an electron in layer $l$ and orbital $o$ with momentum $\mathbf{k}$. In the clean case there is reflection symmetry between the two layers of a bilayer sandwich, such that the non-interacting Hamiltonian can be written in terms of intra-layer $\hat{H}_0^\parallel$ and interlayer blocks $\hat{H}_0^\perp$
\begin{equation}
\hat{H}_0(\mathbf{k}) =
    \begin{pmatrix}
		\hat{H}_0^{\parallel}(\mathbf{k})	
        & \hat{H}_0^{\perp}(\mathbf{k}) e^{ik_zd} \\
		\hat{H}_0^{\perp}(\mathbf{k}) e^{-ik_zd}
        & \hat{H}_0^{\parallel}(\mathbf{k})	
	\end{pmatrix},
 \label{eq:bilayerHamiltonian}
\end{equation}
where the hats refer to the matrices in orbital space. In the following, we drop explicit momentum and frequency dependence and exchange subscripts and superscripts whenever it is convenient for readability. The phase factors $e^{\pm ik_z d}$ arise from the Fourier transform for a bilayer sandwich with thickness $d$. The presence of reflection symmetry between the two layers allows to diagonalize the non-interacting Hamiltonian by the unitary transformations towards bonding-antibonding space (ba-space)
\begin{align}
\hat{V} = \frac{1}{\sqrt{2}}
    \begin{pmatrix}
		\hat{\mathbb{1}}	
        &   \hat{\mathbb{1}} e^{ik_zd} \\
		 \hat{\mathbb{1}} e^{-ik_zd}
        & -\hat{\mathbb{1}}
	\end{pmatrix},
 \ \ \ \
 \hat{H}^{b/a}_{0} =  \hat{H}^{\parallel}_{0} \pm \hat{H}^{\perp}_{0}.
 \label{eq:abTrafo}
\end{align}
Here, the phase factors have been factored out and are no longer present in the ba-space. However, they are important when considering two particle correlation functions \cite{Boetzel2024}.

We next introduce superconductivity on a mean-field level. After applying mean-field approximation, the interaction part reads
\begin{equation}
   \mathcal{H}^{\text{MF}}_{\text{int}} = \sum_{\mathbf{k}}\sum_{l_1,l_2}\sum_{o_1,o_2} \hat{\mathcal{D}}_{l_1o_1;l_2o_2}(\mathbf{k}) c^{\dagger}_{l_1o_1\uparrow}(\mathbf{k})c^{\dagger}_{l_2o_2\downarrow}(-\mathbf{k}) + h.c.,  
\end{equation}
and we can again introduce the ba-block structure
\begin{equation}
\hat{\mathcal{D}}(\mathbf{k}) =
    \begin{pmatrix}
		\hat{\Delta}^{\parallel}(\mathbf{k})	
        & \hat{\Delta}^{\perp}(\mathbf{k}) e^{ik_zd} \\
		\hat{\Delta}^{\perp}(\mathbf{k}) e^{-ik_zd}
        & \hat{\Delta}^{\parallel}(\mathbf{k})	
	\end{pmatrix},
 \label{eq:bilayerHamiltonianMFSC}
\end{equation}
which is again diagonalized using the ba-transformation $\hat{V}$. The superconducting gaps in ba-space are $\hat{\Delta}^{b/a} = \hat{\Delta}^\parallel \pm \hat{\Delta}^{\perp}$.

To simulate the effect of the electron irradiation, we consider randomly distributed point-like impurities at Ni positions in both NiO$_2$ layers. Such impurities locally break the reflection symmetry and consequently mix bonding and antibonding blocks. Impurity averaging will re-introduce the ba-block structure, but one has to be careful to perform the averaging in both layers separately \cite{Hofmann1990}. The impurity matrices for the upper and lower layer read
\begin{equation}
\hat{W}_{1} =     
    \begin{pmatrix}
		\hat{W}_{\mathbf{k}o,\mathbf{k}'o'}  
        & 0 \\
		0	
        & 0
	\end{pmatrix}\otimes \hat{\tau}_3,
\hat{W}_{2} =     
    \begin{pmatrix}
		  0
        & 0 \\
		0	
        & \hat{W}_{\mathbf{k}o,\mathbf{k}'o'}
	\end{pmatrix}\otimes\hat{\tau}_3,
\end{equation}
where $\hat{\tau}_i$ denotes the $i-$th Pauli matrix in the Gor'kov-Nambu space. We consider the self-energy arising due to impurity scattering in the non-crossing approximation and assume the same impurity concentration in both layers. For simplicity, we focus on the intraorbital scattering, which preserves $C_4$ symmetry, i.e. $\hat{W}_{\mathbf{k}o,\mathbf{k}'o'} = W \hat{\mathbb{1}}$. After impurity averaging, we obtain the self-energy in the form
\begin{align}
  \hat{\Sigma} &= n_{\text{imp}} \sum_{\mathbf{k}} \sum_{l=1}^{2} \hat{W}_{l} \hat{G}_\mathbf{k} \hat{W}_{l} \nonumber \\
&= n_{\text{imp}} W^2 \sum_{\mathbf{k}}  (\hat{\tau}_3 \hat{G}^\parallel_\mathbf{k} \hat{\tau}_3) \otimes \mathbb{1}_{\text{layer}}.
\label{Eq:GeneralSelfEnergy}
\end{align}
Importantly, the above expression corresponds to an averaging over bonding and antibonding blocks. To see this, the relation $2 \hat{G}^\parallel_\mathbf{k} = \hat{G}^b_\mathbf{k} + \hat{G}^a_\mathbf{k}$ for the Green's functions can be inserted, which can again be shown by applying the ba-transformation. This ba-averaging is the direct consequence of the local breaking of reflection symmetry. Note that an impurity matrix of the form $\hat{W}_1 + \hat{W}_2$ is not creating such an averaging \cite{Hofmann1990}. The Green's functions are self-consistently calculated via
\begin{equation}
    \hat{G}_{b/a}^{-1}(k) = i\omega_n \hat{\mathbb{1}} -\hat{H}^{b/a}_{0}(\mathbf{k})\tau_3 - \hat{\Delta}^{b/a}(\mathbf{k})\tau_1  - \hat{\Sigma}(i\omega_n).
    \label{Eq:baGF}
\end{equation}
Note, the self-energy is unity in the layer space and, therefore, transforms trivially to ba-space, which results from impurity averaging restoring the global reflection symmetry.

\section{Single-orbital model}

%%%%%%%%%%%%%%%%%%%%%%%%%%%%%%%%%%%%%%%%%%%%%%%%%%%%%%%%%%%%%
\begin{figure*}[t]
      \includegraphics[width=\linewidth]{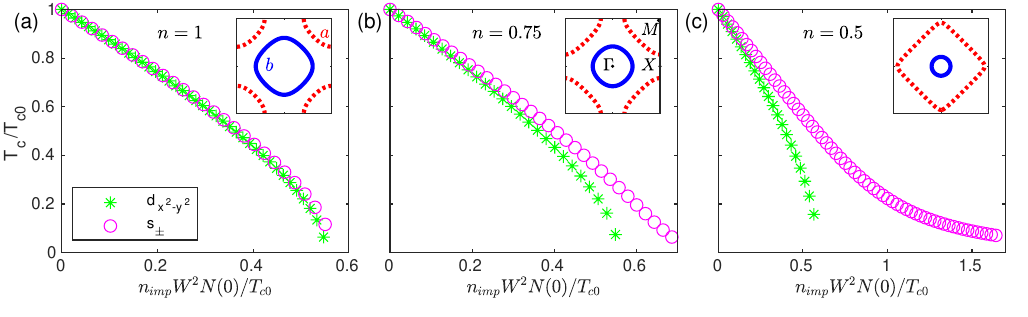}
	\caption{(color online) Calculated normalized $T_c/T_{c0}$ curves as a function of impurity concentrations for bonding-antibonding $s_{\pm}$-wave ($\Delta_{b/a} = \pm \Delta_\perp$) and $d$-wave ($\Delta_{b/a} = \Delta_d \gamma_d$) Cooper-pairings for a bilayer model with single orbital for various band fillings. The inset shows the corresponding Fermi surface with bonding and antibonding bands in blue and red, respectively. The imbalance (here it is the particle-hole asymmetry) between bonding and antibonding bands increases from left to right. The interaction $J$ and $t_\perp$ are modified such that $T_{c0}$ is roughly the same for all cases. $N(0)$ in the x-label denotes the density of states in the normal state at the Fermi level.} \label{fig1}
\end{figure*}	
%%%%%%%%%%%%%%%%%%%%%%%%%%%%%%%%%%%%%%%%%%%%%%%%%%%%%%%%%%%%%%

To study the effect of the averaging over ba-space on the $T_c$-suppression, it is instructive to consider first a simple bilayer model with only one orbital, considered e.g. in Ref.~\cite{Maier2011}. To be specific, we employ a $t-J$-like  bilayer model with in-plane nearest neighbor hopping $t$ and interlayer hopping $t_\perp$ 
\begin{align}
\mathcal{H}_0 = - t \sum_{\langle i, j\rangle,\sigma}\sum_{r=1}^2 c_{i,r,\sigma}^{\dagger} c_{j,r,\sigma} - t_{\perp} \sum_{i,\sigma} (c_{i,1,\sigma}^{\dagger} c_{i,2,\sigma}+\text{H.c.}).
\end{align}
Similarly, we consider the superexchange in-plane interaction $J$ and inter-layer interaction $J_\perp$
\begin{align}
\mathcal{H}_{\text{int}}=   J \sum_{\langle i, j\rangle}\sum_{r=1}^2
\mathbf{S}_{i,r} . \mathbf{S}_{j,r}
 +J_{\perp} \sum_{i} \mathbf{S}_{i,1} . \mathbf{S}_{i,2}.  
\end{align}
The summation brackets $\langle i, j\rangle$ indicate summation over nearest neighbors only and $r$ is the layer index.  Unless stated otherwise, we assume $J_\perp =  (t_\perp/t)^2 J$, which is expected relation once the $t-J$ model is derived from a Hubbard-type model in the strong-coupling limit. The non-interacting parts yields ba-dispersions $\epsilon^{b/a}_\mathbf{k} = -2t (\cos(k_x)+\cos(k_y) \pm t_\perp - \mu$. We perform a mean-field decoupling from which we obtain the following gap equations (neglecting possible in-plane triplet part):
 \begin{align}
\Delta_{d/s} = -VT \sum_{n,\mathbf{k}} \gamma_{d/s}  \frac{\tilde{\Delta}_{\mathbf{k},b}}{\tilde{\omega}_{n,b}^{2}+\tilde{\epsilon}_{\mathbf{k},b}^{2}+\tilde{\Delta}_{\mathbf{k},b}^{2}} + (b \leftrightarrow a)  \\
\Delta_{\perp} = -V_\perp T \sum_{n,\mathbf{k}}  \frac{\tilde{\Delta}_{\mathbf{k},b}}{\tilde{\omega}_{n,b}^{2}+\tilde{\epsilon}_{\mathbf{k},b}^{2}+\tilde{\Delta}_{\mathbf{k},b}^{2}} - (b \leftrightarrow a) 
\label{Eq:GapPerp}
\end{align}
with $\gamma_{d/s} = (\cos(k_x) \mp \cos(k_y))/2$ and $\Delta_{b/a} = \gamma_s \Delta_s + \gamma_d \Delta_d \pm \Delta_\perp$ and $V = -3 J$ and $V_\perp = -(3/8) J_\perp$. Here, $\Delta_{s/d}$ denotes the in-plane extended $s$-wave and $d_{x^2-y^2}$-wave gap functions, respectively, whereas $\Delta_\perp$ is the interlayer $s$-wave gap and in the case of opposite signs between bonding and antibonding bands refers to the bonding-antibonding $s_\pm$-wave solution. The $d_{xy}$-wave solution would have a $\sin k_x \sin k_y$ functional form in the plane. We do not consider it specifically as it behaves similarly to the $d_{x^2-y^2}-$wave solution in the presence of the non-magnetic point-like impurities. In addition, the $d_{x^2-y^2}-$wave solution can also have an interlayer component $ \pm \Delta^d_{\perp} \gamma_d$ and yield either in-phase or out-of-phase locking between two neighboring planes. At the same time, its magnitude is expected to be much smaller than that in the $s$-wave case, since the $d$-wave solution is only dominant for the smaller values of the interlayer hopping. For simplicity, we assume it to be small and consider the in-phase locking of the $d$-wave solution between the layers.

In the clean case, the $d_{x^2-y^2}$-wave solution for the given model is found at half-filling for $t_\perp \lesssim 1.1 t$. For larger $t_\perp$, the interlayer ba-$s_\pm$ solution becomes dominant at half-filling. Importantly, as $\Delta_s$ and $\Delta_\perp$ correspond to gap structures of $A_{1g}$ symmetry, they always occur simultaneously. In the following, we will focus on the pure interlayer ba-$s_\pm$ solution and the mixed solution is discussed in the Supplementary Material. In this one-orbital model, the bonding and antibonding Green's functions in Eq.~(\ref{Eq:baGF}) are given by 
\begin{equation}
    \hat{G}_{b/a}^{-1} = (i\omega_n - \Sigma_0) \hat{\tau}_0  - (\epsilon^{b/a}_\mathbf{k}+\Sigma_1) \hat{\tau}_3 - (\Delta^{b/a}_{\mathbf{k}}  + \Sigma_1)\hat{\tau}_1 
    \label{Eq_baGFToyModel}
\end{equation}
and the renormalized quantities $i\tilde{\omega}_n = i\omega_n - \Sigma_0$, $\tilde{\epsilon}^{b/a}_{\mathbf{k}} = \epsilon^{b/a}_\mathbf{k} + \Sigma_3$ and $\tilde{\Delta}^{b/a}_{\mathbf{k}} = \Delta^{b/a}_{\mathbf{k}} + \Sigma_1$ are defined as
\begin{align}
    \Sigma_0 &=\frac{1}{2}  i\tilde{\omega}_n n_{\text{imp}} W^2  \sum_{\mathbf{k}} \frac{1}{\tilde{\omega}_{n,b}^{2}+\tilde{\epsilon}_{\mathbf{k},b}^{2}+\tilde{\Delta}_{\mathbf{k},b}^{2}} + (b \leftrightarrow a) \\
    \Sigma_3 &= \frac{1}{2} n_{\text{imp}} W^2  \sum_{\mathbf{k}} \frac{\tilde{\epsilon}^{b}_\mathbf{k}}{\tilde{\omega}_{n,b}^{2}+\tilde{\epsilon}_{\mathbf{k},b}^{2}+\tilde{\Delta}_{\mathbf{k},b}^{2}} + (b \leftrightarrow a) \\
    \Sigma_1 &= \frac{1}{2} n_{\text{imp}} W^2  \sum_{\mathbf{k}} \frac{\tilde{\Delta}_{b,\mathbf{k}}}{\tilde{\omega}_{n,b}^{2}+\tilde{\epsilon}_{\mathbf{k},b}^{2}+\tilde{\Delta}_{\mathbf{k},b}^{2}} + (b \leftrightarrow a)
    \label{Eq:AnomalousSelfEnergyToyModel}
\end{align}
Recall, that for the $d_{x^2-y^2}$-wave gap function, $\Sigma_1$ vanishes by symmetry in momentum space as can be readily seen from inserting $\Delta^{b/a}_\mathbf{k} = \gamma_d \Delta_d$ in Eq.~(\ref{Eq:AnomalousSelfEnergyToyModel}). The interlayer ba-$s_{\pm}$-wave solution with $\Delta_{b/a} = \pm \Delta_\perp$ has no dependence on momentum $\mathbf{k}$. As we are interested in the evolution of the superconducting transition temperature, $T_c$, when the superconducting order parameter approaches the limit $\Delta \rightarrow 0$, we rewrite Eq.~(\ref{Eq:AnomalousSelfEnergyToyModel}) as
\begin{align}
    \Sigma_1 &= \frac{1}{2} n_{\text{imp}} W^{2}  \left[ \tilde{\Delta}_{b} \sum_{\mathbf{k}}  \frac{   1 }{\tilde{\omega}_{n}^{2}+\tilde{\epsilon}_{\mathbf{k},b}^{2}} +  (b \leftrightarrow a)
    \right] \nonumber \\
    & =: \tilde{\Delta}_{b} \Omega_{b} + \tilde{\Delta}_{a} \Omega_{a}.
    \label{Eq:selfEnergybaTcLimit}
\end{align}
and obtain two equations $\tilde{\Delta}_{b/a} = \Delta_{b/a} + \tilde{\Delta}_{b} \Omega_{b} + \tilde{\Delta}_{a} \Omega_{a}$. A similar set of equations was discussed in the context of iron-based superconductors \cite{Efremov2011,Korshunov2016} with one important difference. In particular, here the inter-band coefficients are not symmetric because in general $\Omega_{a} \neq \Omega_b$. Solving for $\tilde{\Delta}_{b/a}$, we find
\begin{equation}
     \tilde{\Delta}_{b/a} = \frac{1}{1-\Omega_{b}-\Omega_{a}} \left[ {\Delta}_{b/a} + \Omega_{a/b} ({\Delta}_{a/b}-{\Delta}_{b/a}) \right]
\end{equation}
and after inserting $\Delta_{b/a} = \pm \Delta_\perp$ explicitly, we obtain
\begin{align}
    \tilde{\Delta}_{b/a} = \pm \Delta_\perp \left( 1 + \frac{\Omega_{b/a} - \Omega_{a/b}}{1-\Omega_{b}-\Omega_{a}} \right). 
    \label{Eq:BAGapCorrection}
\end{align}
As follows from Eq.~(\ref{Eq:BAGapCorrection}), the $T_c$ suppression for the ba-$s_{\pm}$-wave depends on the difference between $\Omega_{b}$ and $\Omega_{a}$. For the specific case $\Omega_{b} = \Omega_{a}$, we find $\tilde{\Delta}_{b/a} = \pm \Delta_\perp$ and this version of the ba-$s_{\pm}$-wave solution will be suppressed by potential point-like impurities in the same way as the $d$-wave superconducting state with $\Sigma_1 = 0$. However, if there is an imbalance (in this particular case it is particle-hole asymmetry) between bonding and antibonding sub-spaces, there will larger deviation from the AG behavior. Moreover, finite $\Omega_{b} \neq \Omega_{a}$ induces different magnitudes for the bonding and superconducting gap functions $\tilde{\Delta}_{b/a}$ as the second term on the r.h.s. of Eq.~(\ref{Eq:BAGapCorrection}) is of different sign. In particular, for increasing impurity density, $\Omega_{b} + \Omega_{a}$ approaches unity and $\tilde{\Delta}_{b/a}$ must change its sign, if $\Omega_{a/b} > \Omega_{b/a} $, respectively. Therefore, a $s_{\pm}$-wave to $s_{++}$-wave crossover happens, which already was discussed previously in the context of iron-based superconductors \cite{Efremov2011,Korshunov2016}. As $\Omega_{b} + \Omega_{a}$ approaches unity, one might be worried about a divergence due to the denominator. However, this is of course not the case because of the concurrent suppression of $\Delta_\perp$. Note that this can be understood with the renormalized Matsubara frequency $\tilde{\omega}_{n} = \omega_n/(1-\Omega_{b}-\Omega_{a})$ entering the gap equation  Eq.~(\ref{Eq:GapPerp}). The renormalization of the quasiparticle energies is typically negligible. Note that the numerator in the correction term in Eq.~(\ref{Eq:BAGapCorrection}) depends linearly on the impurity density. Consequently, the deviation from AG behavior is small for low impurity densities but sizable for higher impurity densities. 

For our simple model, we employ $\epsilon^{b/a}_\mathbf{k} = -2t (\cos(k_x) + \cos(k_y)) \pm t_\perp - \mu$. At half-filling, $\mu = 0$, and the relation $\epsilon^{b}_{\mathbf{k}} = -\epsilon^{a}_{\mathbf{k}+(\pi,\pi)}$ guarantees $\Omega_{b} = \Omega_{a}$ (note that for this case particle-hole symmetry is present). Therefore, moving away from half-filling allows us to tune the particle-holy asymmetry between bonding and antibonding bands and, consequently, between $\Omega_{b}$ and $\Omega_{a}$. In Fig.~\ref{fig1} the $T_c$ suppression for exemplary band fillings $n = 1$, $n = 0.75$ and $n=0.5$ are compared for the pure interlayer ba-$s_{\pm}$-wave and $d_{x^2-y^2}$-wave Cooper-pairing scenarios. Clearly, the qualitative behavior is changing as the imbalance between bonding and antibonding bands increases. At half-filling, both scenarios give the same qualitative AG shape (Fig.~\ref{fig1}(a)). Upon lowering the filling to $n=0.75$, the $T_c$ curve for the interlayer $s_{\pm}$-wave starts to deviate from AG form close to the critical impurity density at which superconductivity is completely suppressed (Fig.~\ref{fig1}(b)). At quarter-filling, $n=0.5$, the initial slope is still comparable for both scenarios but the $T_c$ curve for the interlayer ba-$s_{\pm}$-wave state changes from convex to concave shape and persists up to more than three times larger impurity densities. We adopt $J$ and $t_\perp$ such that we have roughly the same $T_c$ in the clean state. The results are not sensitive to the variation of $t_\perp$. Fixing $t_\perp$ and varying $J$ and $J_\perp$ independently gives qualitatively identical results as shown in Fig.~S4 of the Supplemental Material \cite{Supplement}. More details on the numerical calculation can also be found in the Supplemental Material.

\section{Multiorbital model of La-327}

%%%%%%%%%%%%%%%%%%%%%%%%%%%%%%%%%%%%%%%%%%%%%%%%%%%%%%%%%%%%%
\begin{figure}[]
      \includegraphics[width=\columnwidth]{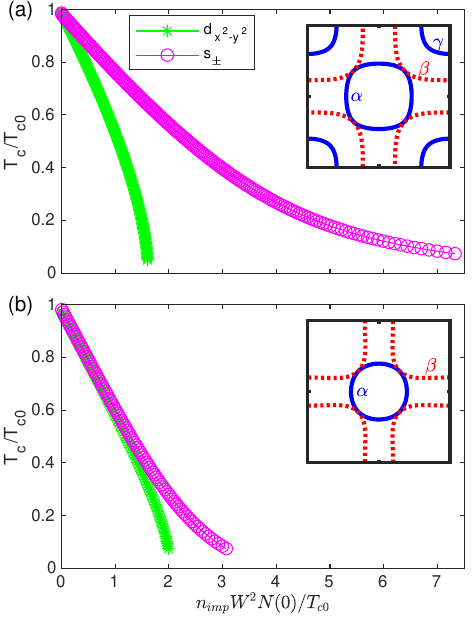}
	\caption{(color online) Calculated normalized $T_c/T_{c0}$ suppression curves by non-magnetic impurities as a function of impurity concentrations  for La-327~\cite{LuoModel,lechermann2023electronic} for the interlayer ba-$s_{\pm}$-wave $\Delta_{b/a} = \pm \Delta_\perp$ and intralayer $d$-wave ($\Delta_{b/a} = \Delta_d \gamma_d$) order parameters. The inset shows the corresponding Fermi surface where the blue and red colors refer to bonding and antibonding Fermi surface sheets, respectively. The interaction $J$ and $J_\perp$ are chosen such that $T_c \approx 80$ K. In (b) the on-site orbital energies are chosen such that the $\gamma$ pocket is shifted 50 meV below the Fermi surface as it is the case for the La-327 at ambient pressure~\cite{yang2024orbital,li2024electronic}, keeping the band filling unchanged.} \label{fig2}
\end{figure}	
%%%%%%%%%%%%%%%%%%%%%%%%%%%%%%%%%%%%%%%%%%%%%%%%%%%%%%%%%%%%%%

We now turn to the analysis of the $T_c$ suppression in pressurized La-327 based on a two-orbital model Hamiltonian. The non-interacting part is described by the tight-binding parametrization, based on $e_g$-manifold consisting of Ni 3$d_{x^2-y^2}$ and 3$d_{z^2}$ orbitals in each layer, taken from Ref.~\cite{LuoModel}, yet another typical tight-binding parametrization with larger set of hoppings \cite{lechermann2023electronic} would give similar results. Importantly, the interlayer intraorbital hopping between $d_{z^2}$ orbitals is larger than the intralayer intraorbital hopping between $d_{x^2-y^2}$ orbitals by a ratio of about 1.3. The corresponding Fermi surface is shown in the inset of Fig.~\ref{fig2}(a). In the context of our discussion so far, it is important to note that there is a large particle-hole asymmetry between bonding and antibonding bands as there are two bonding ($\alpha$ and $\gamma$) pockets, whereas there is only a single antibonding ($\beta$) pocket. While the $\alpha$ and $\beta$-bands are due to bonding-antibonding splitting of mostly 3$d_{x^2-y^2}$-orbitals, the bonding 3$d_{z^2}$ states give rise to the flattish $\gamma$ pocket within DFT results \cite{sun23} and the antibonding  3$d_{z^2}$-states are well above the Fermi level. At the same time, low-temperature ARPES experiments at ambient pressure do not observe the $\gamma$ pocket, contrary to the DFT prediction \cite{yang2024orbital,li2024electronic}. One explanation for
this discrepancy is linked to a spin-density-wave (SDW) ordering transition, taking place around $T_{\rm SDW}\sim 150$\,K~\cite{327sdw2024} and possibly splitting off the $\gamma$-pocket from the Fermi level~\cite{lechermann24}. In any case, it leaves the pocket's fate in the high-pressure phase open to the effect of strong electronic correlation \cite{lechermann2023electronic,wang2024electronicmagneticstructuresbilayer,ryee2023critical}. To mimic the effect of an absent $\gamma$ pocket we modified the onsite energies of the two-orbital model to shift the $\gamma$ band below $E_F$ by about 50 meV keeping the overall band filling unchanged.

The interactions are included on the level of superexchange-like terms, similar to Ref.~\cite{luo2024high} 
\begin{align}
%\mathcal{H}_0 = &  \sum_{i_1, i_2,s}\sum_{r_1,r_2}^2\sum_{o_1,o_2} \left(-t^{i_1,i_2}_{r_1,o_1;r_2,o_2} c_{i_1,r_1,o_1,s}^{\dagger} c_{i_2,r_2,o_2,s}+H . c .\right) \\
\mathcal{H}_{int} = &J \sum_{\langle i_1, i_2\rangle}\sum_{r=1}^2 \left(\mathbf{S}_{i_1,r,x} . \mathbf{S}_{i_2,r,x} \right) + J_{\perp} \sum_{i}\left(\mathbf{S}_{i,1,z} . \mathbf{S}_{i,2,z}\right), 
\end{align}
and restricted to in-plane nearest neighbor interaction $J$ between $d_{x^2-y^2}$ orbitals and the interlayer ones within a bilayer sandwich between $d_{z^2}$ orbitals $J_\perp$ \cite{luo2024high}. However, note that the different orbitals are well hybridized, which guarantees strong coupling of the planes also for the $d_{x^2-y^2}$-orbitals. 
In particular, the hybridization term between $d_{x^2-y^2}$ and $d_{z^2}$ orbitals follows the momentum dependence of $\gamma_d$. The consequence is that a $d_{x^2-y^2}$-wave superconducting gap function with an intraorbital gap $\Delta^{b/a}_{x^2-y^2}$ also induces an interorbital gap $\Delta^{b/a}_{\text{inter}}$ following $\gamma_d^2$ (see Fig.~S2 \cite{Supplement} for illustration). Hence, one obtains with orbital hybridization a non-vanishing anomalous self-energy $\Sigma_1$ even for the $d$-wave case. However, there is no strong deviation from AG-behavior for La-327 as shown in Fig.~S1 \cite{Supplement}. For the interlayer ba-$s_\pm$-wave scenario with constant intraorbital $\Delta^{b/a}_{z^2}$ the hybridization forces $\Delta^{b/a}_{\text{inter}}$ to follow the form factor $\gamma_d$. 

Having several orbitals brings additional complexity and therefore we proceed numerically. In particular, we solve Eq.~(\ref{Eq:GeneralSelfEnergy}) and Eq.~(\ref{Eq:baGF}) and the superconducting gap equations
\begin{align}
\Delta_{d,s} = -VT \sum_{n,\mathbf{k}} \gamma_{d,s}  \left( F^b_{x^2-y^2} + F^a_{x^2-y^2} \right) \\
\Delta_{\perp} = -V_\perp T \sum_{n,\mathbf{k}}  \left( F^b_{z^2} - F^a_{z^2} \right)
\end{align}
self-consistently where $F^{b/a}_{i}$ denotes the $i$-th intraorbital component of the anomalous ba-Green's function. To find $T_c$ we solve the gap equations self-consistently upon varying the impurity concentrations until the gap value converges to a value $\leq 10^{-5}$ eV and further restrict the momentum summation to a 100 meV shell around the Fermi level. To keep the tetragonal symmetry we also assume that the impurity scattering is diagonal in the orbital space. 

The resulting $T_c$ suppression with the $d_{x^2-y^2}$-wave and interlayer ba-$s_\pm$-wave solutions are plotted in Fig.~\ref{fig2} with (a) and without (b) the $\gamma$-pocket being at the Fermi level. We find that for the electronic structure of La-327 there is a significant difference of the T$_c$ curves, produced by point-like non-magnetic impurities. While $d-$wave superconductivity is suppressed following more or less the conventional AG pair-breaking behavior, expected for the $d-$wave superconducting state, and the difference arises due to multi-orbital character of superconductivity, the bonding-antibonding $s_{\pm}$-wave superconducting state deviates strongly. Although $T_c$ is reduced for increasing impurity concentration, its value finally saturates due to $s_{\pm}$-wave to $s_{++}$-wave crossover, and the latter is robust to the non-magnetic impurity scattering. 

Furthermore, changing the fermiology of the initial model from having the low-energy regime dominated by the bonding subspace to having a more balanced appearance of both ba-subspace, i.e. without the bonding $z$-electronic states ($\gamma$-pocket) at the Fermi level, shifts the crossover from convex to concave shape of the $T_c$ curve as a function of impurity concentration. To be more precise, we even find a change in dominance from bonding to antibonding subspace near the Fermi level as apparent from the inset in Fig.~\ref{fig2}(b). Therefore, the $s_{++}$-wave at higher impurity densities will adopt the sign of the antibonding $\beta$ band. This implies that for some position of the $\gamma$ pocket near 50 meV below the Fermi surface, bonding and antibonding subspaces must be balanced in the low-energy regime. In a small region around this position, the interlayer ba-$s$-wave and $d$-wave $T_c$ suppression curves will not be distinguishable. For this rather unlikely case, together with other experimental data providing evidence for the interlayer ba-$s$-wave, making a precise comparison to the future experimental data would allow to make indirect conclusions on the fate of the $\gamma$-pocket in this case and finally its role for superconductivity in La-327. As soon as the $\gamma$ pocket is shifted below the Fermi level, the dominant superconducting gap structure changes, making the interlayer ba-$s_\pm$-wave not the leading solution and it requires much higher interaction $J_\perp$ to stabilize \cite{luo2024high,Supplement}. 
%Furthermore, the $\gamma$ pocket is not participating in the $d$-wave superconducting gap and fewer states participate in superconductivity, which is the reason for the different initial slopes in Fig.~\ref{fig2}(a).  

\section{Summary}

To summarize, we analyze the suppression of unconventional superconductivity in pressurized La-327 by non-magnetic point-like impurities assuming sign-changing bonding-antibonding $s_{\pm}$-wave and $d$-wave symmetries of the superconducting order parameters, currently discussed in the literature.  We show that each superconducting order parameter is suppressed in a different fashion. While the suppression of the $d$-wave superconductivity follows more or less the Abrikosov-Gor'kov characteristic behavior for the unconventional superconducting order, the bonding-antibonding sign-changing $s_{\pm}$-wave superconductivity is suppressed differently, depending on the asymmetry of the bonding-antibonding electronic bands. A weak asymmetry of the bonding-antibonding bands with respect to their location at the Fermi level would result in the suppression of the $s_{\pm}$-wave superconducting state similar to the $d$-wave case. Away from this high-symmetry point of the bilayer model, $s_{\pm}$-wave superconductivity is more robust against non-magnetic impurity scattering and the $T_c$ curve experiences a crossover from convex to concave shape, which signals the corresponding evolution from $s_{\pm}$-wave to $s_{++}$-wave superconductivity. This characteristic behavior makes it an indicative test for experimental verification by studying $T_c$ evolution as a function of electron irradiation. Furthermore, the robustness of superconductivity in the $s_{\pm}$-wave case also depends on the bonding-antibonding band structure asymmetry with respect to the Fermi level of the normal state model and could be used to elucidate the role of the $\gamma$-pocket in the pressurized La-327 samples. Another interesting test would be to study the change of the slope of the upper critical field under the effect of electron irradiation like it was proposed for the iron-based superconductors \cite{Kogan2023}.

\section{Acknowledgements} 
The work is supported by the German Research Foundation within the bilateral
NSFC-DFG Project ER 463/14-1, and by Grant-in-Aid for Scientific Research (KAKENHI) (No. JP22H00105) from Japan Society for the Promotion of Science (JSPS). 

\bibliography{literatur}

\begin{thebibliography}{58}%
\makeatletter
\providecommand \@ifxundefined [1]{%
 \@ifx{#1\undefined}
}%
\providecommand \@ifnum [1]{%
 \ifnum #1\expandafter \@firstoftwo
 \else \expandafter \@secondoftwo
 \fi
}%
\providecommand \@ifx [1]{%
 \ifx #1\expandafter \@firstoftwo
 \else \expandafter \@secondoftwo
 \fi
}%
\providecommand \natexlab [1]{#1}%
\providecommand \enquote  [1]{``#1''}%
\providecommand \bibnamefont  [1]{#1}%
\providecommand \bibfnamefont [1]{#1}%
\providecommand \citenamefont [1]{#1}%
\providecommand \href@noop [0]{\@secondoftwo}%
\providecommand \href [0]{\begingroup \@sanitize@url \@href}%
\providecommand \@href[1]{\@@startlink{#1}\@@href}%
\providecommand \@@href[1]{\endgroup#1\@@endlink}%
\providecommand \@sanitize@url [0]{\catcode `\\12\catcode `\$12\catcode `\&12\catcode `\#12\catcode `\^12\catcode `\_12\catcode `\%12\relax}%
\providecommand \@@startlink[1]{}%
\providecommand \@@endlink[0]{}%
\providecommand \url  [0]{\begingroup\@sanitize@url \@url }%
\providecommand \@url [1]{\endgroup\@href {#1}{\urlprefix }}%
\providecommand \urlprefix  [0]{URL }%
\providecommand \Eprint [0]{\href }%
\providecommand \doibase [0]{https://doi.org/}%
\providecommand \selectlanguage [0]{\@gobble}%
\providecommand \bibinfo  [0]{\@secondoftwo}%
\providecommand \bibfield  [0]{\@secondoftwo}%
\providecommand \translation [1]{[#1]}%
\providecommand \BibitemOpen [0]{}%
\providecommand \bibitemStop [0]{}%
\providecommand \bibitemNoStop [0]{.\EOS\space}%
\providecommand \EOS [0]{\spacefactor3000\relax}%
\providecommand \BibitemShut  [1]{\csname bibitem#1\endcsname}%
\let\auto@bib@innerbib\@empty
%</preamble>
\bibitem [{\citenamefont {Sun}\ \emph {et~al.}(2023)\citenamefont {Sun}, \citenamefont {Huo}, \citenamefont {Hu}, \citenamefont {Li}, \citenamefont {Han}, \citenamefont {Tang}, \citenamefont {Mao}, \citenamefont {Yang}, \citenamefont {Wang}, \citenamefont {Cheng}, \citenamefont {Yao}, \citenamefont {Zhang},\ and\ \citenamefont {Wang}}]{sun23}%
  \BibitemOpen
  \bibfield  {author} {\bibinfo {author} {\bibfnamefont {H.}~\bibnamefont {Sun}}, \bibinfo {author} {\bibfnamefont {M.}~\bibnamefont {Huo}}, \bibinfo {author} {\bibfnamefont {X.}~\bibnamefont {Hu}}, \bibinfo {author} {\bibfnamefont {J.}~\bibnamefont {Li}}, \bibinfo {author} {\bibfnamefont {Y.}~\bibnamefont {Han}}, \bibinfo {author} {\bibfnamefont {L.}~\bibnamefont {Tang}}, \bibinfo {author} {\bibfnamefont {Z.}~\bibnamefont {Mao}}, \bibinfo {author} {\bibfnamefont {P.}~\bibnamefont {Yang}}, \bibinfo {author} {\bibfnamefont {B.}~\bibnamefont {Wang}}, \bibinfo {author} {\bibfnamefont {J.}~\bibnamefont {Cheng}}, \bibinfo {author} {\bibfnamefont {D.-X.}\ \bibnamefont {Yao}}, \bibinfo {author} {\bibfnamefont {G.-M.}\ \bibnamefont {Zhang}},\ and\ \bibinfo {author} {\bibfnamefont {M.}~\bibnamefont {Wang}},\ }\href {https://doi.org/10.1038/s41586-023-06408-7} {\bibfield  {journal} {\bibinfo  {journal} {Nature}\ }\textbf {\bibinfo {volume} {621}},\ \bibinfo {pages} {493} (\bibinfo {year} {2023})}\BibitemShut {NoStop}%
\bibitem [{\citenamefont {Hou}\ \emph {et~al.}(2023)\citenamefont {Hou}, \citenamefont {Yang}, \citenamefont {Liu}, \citenamefont {Li}, \citenamefont {Shan}, \citenamefont {Ma}, \citenamefont {Wang}, \citenamefont {Wang}, \citenamefont {Guo}, \citenamefont {Sun}, \citenamefont {Uwatoko}, \citenamefont {Wang}, \citenamefont {Zhang}, \citenamefont {Wang},\ and\ \citenamefont {Cheng}}]{JunHou:117302}%
  \BibitemOpen
  \bibfield  {author} {\bibinfo {author} {\bibfnamefont {J.}~\bibnamefont {Hou}}, \bibinfo {author} {\bibfnamefont {P.-T.}\ \bibnamefont {Yang}}, \bibinfo {author} {\bibfnamefont {Z.-Y.}\ \bibnamefont {Liu}}, \bibinfo {author} {\bibfnamefont {J.-Y.}\ \bibnamefont {Li}}, \bibinfo {author} {\bibfnamefont {P.-F.}\ \bibnamefont {Shan}}, \bibinfo {author} {\bibfnamefont {L.}~\bibnamefont {Ma}}, \bibinfo {author} {\bibfnamefont {G.}~\bibnamefont {Wang}}, \bibinfo {author} {\bibfnamefont {N.-N.}\ \bibnamefont {Wang}}, \bibinfo {author} {\bibfnamefont {H.-Z.}\ \bibnamefont {Guo}}, \bibinfo {author} {\bibfnamefont {J.-P.}\ \bibnamefont {Sun}}, \bibinfo {author} {\bibfnamefont {Y.}~\bibnamefont {Uwatoko}}, \bibinfo {author} {\bibfnamefont {M.}~\bibnamefont {Wang}}, \bibinfo {author} {\bibfnamefont {G.-M.}\ \bibnamefont {Zhang}}, \bibinfo {author} {\bibfnamefont {B.-S.}\ \bibnamefont {Wang}},\ and\ \bibinfo {author} {\bibfnamefont {J.-G.}\ \bibnamefont {Cheng}},\ }\bibfield  {title} {\bibinfo {title} {Emergence of
  high-temperature superconducting phase in pressurized {La$_{3}$Ni$_{2}$O$_7$} crystals},\ }\href {https://doi.org/10.1088/0256-307X/40/11/117302} {\bibfield  {journal} {\bibinfo  {journal} {Chinese Physics Letters}\ }\textbf {\bibinfo {volume} {40}},\ \bibinfo {eid} {117302} (\bibinfo {year} {2023})}\BibitemShut {NoStop}%
\bibitem [{\citenamefont {Zhang}\ \emph {et~al.}(2023{\natexlab{a}})\citenamefont {Zhang}, \citenamefont {Su}, \citenamefont {Huang}, \citenamefont {Sun}, \citenamefont {Huo}, \citenamefont {Shan}, \citenamefont {Ye}, \citenamefont {Yang}, \citenamefont {Li}, \citenamefont {Smidman} \emph {et~al.}}]{zhang2023high}%
  \BibitemOpen
  \bibfield  {author} {\bibinfo {author} {\bibfnamefont {Y.}~\bibnamefont {Zhang}}, \bibinfo {author} {\bibfnamefont {D.}~\bibnamefont {Su}}, \bibinfo {author} {\bibfnamefont {Y.}~\bibnamefont {Huang}}, \bibinfo {author} {\bibfnamefont {H.}~\bibnamefont {Sun}}, \bibinfo {author} {\bibfnamefont {M.}~\bibnamefont {Huo}}, \bibinfo {author} {\bibfnamefont {Z.}~\bibnamefont {Shan}}, \bibinfo {author} {\bibfnamefont {K.}~\bibnamefont {Ye}}, \bibinfo {author} {\bibfnamefont {Z.}~\bibnamefont {Yang}}, \bibinfo {author} {\bibfnamefont {R.}~\bibnamefont {Li}}, \bibinfo {author} {\bibfnamefont {M.}~\bibnamefont {Smidman}}, \emph {et~al.},\ }\bibfield  {title} {\bibinfo {title} {High-temperature superconductivity with zero-resistance and strange metal behavior in {La$_{3}$Ni$_{2}$O$_7$}},\ }\href@noop {} {\bibfield  {journal} {\bibinfo  {journal} {arXiv preprint arXiv:2307.14819}\ } (\bibinfo {year} {2023}{\natexlab{a}})}\BibitemShut {NoStop}%
\bibitem [{\citenamefont {Zhou}\ \emph {et~al.}(2023)\citenamefont {Zhou}, \citenamefont {Guo}, \citenamefont {Cai}, \citenamefont {Sun}, \citenamefont {Wang}, \citenamefont {Zhao}, \citenamefont {Han}, \citenamefont {Chen}, \citenamefont {Wu}, \citenamefont {Ding}, \citenamefont {Wang}, \citenamefont {Xiang}, \citenamefont {kwang Mao},\ and\ \citenamefont {Sun}}]{zhou2023evidence}%
  \BibitemOpen
  \bibfield  {author} {\bibinfo {author} {\bibfnamefont {Y.}~\bibnamefont {Zhou}}, \bibinfo {author} {\bibfnamefont {J.}~\bibnamefont {Guo}}, \bibinfo {author} {\bibfnamefont {S.}~\bibnamefont {Cai}}, \bibinfo {author} {\bibfnamefont {H.}~\bibnamefont {Sun}}, \bibinfo {author} {\bibfnamefont {P.}~\bibnamefont {Wang}}, \bibinfo {author} {\bibfnamefont {J.}~\bibnamefont {Zhao}}, \bibinfo {author} {\bibfnamefont {J.}~\bibnamefont {Han}}, \bibinfo {author} {\bibfnamefont {X.}~\bibnamefont {Chen}}, \bibinfo {author} {\bibfnamefont {Q.}~\bibnamefont {Wu}}, \bibinfo {author} {\bibfnamefont {Y.}~\bibnamefont {Ding}}, \bibinfo {author} {\bibfnamefont {M.}~\bibnamefont {Wang}}, \bibinfo {author} {\bibfnamefont {T.}~\bibnamefont {Xiang}}, \bibinfo {author} {\bibfnamefont {H.}~\bibnamefont {kwang Mao}},\ and\ \bibinfo {author} {\bibfnamefont {L.}~\bibnamefont {Sun}},\ }\href@noop {} {\bibinfo {title} {Evidence of filamentary superconductivity in pressurized {La$_{3}$Ni$_{2}$O$_7$} single crystals}} (\bibinfo {year}
  {2023}),\ \Eprint {https://arxiv.org/abs/2311.12361} {arXiv:2311.12361 [cond-mat.supr-con]} \BibitemShut {NoStop}%
\bibitem [{\citenamefont {Zhang}\ \emph {et~al.}(2024{\natexlab{a}})\citenamefont {Zhang}, \citenamefont {Pei}, \citenamefont {Wang}, \citenamefont {Zhao}, \citenamefont {Li}, \citenamefont {Cao}, \citenamefont {Zhu}, \citenamefont {Wu},\ and\ \citenamefont {Qi}}]{zhang2023effects}%
  \BibitemOpen
  \bibfield  {author} {\bibinfo {author} {\bibfnamefont {M.}~\bibnamefont {Zhang}}, \bibinfo {author} {\bibfnamefont {C.}~\bibnamefont {Pei}}, \bibinfo {author} {\bibfnamefont {Q.}~\bibnamefont {Wang}}, \bibinfo {author} {\bibfnamefont {Y.}~\bibnamefont {Zhao}}, \bibinfo {author} {\bibfnamefont {C.}~\bibnamefont {Li}}, \bibinfo {author} {\bibfnamefont {W.}~\bibnamefont {Cao}}, \bibinfo {author} {\bibfnamefont {S.}~\bibnamefont {Zhu}}, \bibinfo {author} {\bibfnamefont {J.}~\bibnamefont {Wu}},\ and\ \bibinfo {author} {\bibfnamefont {Y.}~\bibnamefont {Qi}},\ }\bibfield  {title} {\bibinfo {title} {{Effects of pressure and doping on Ruddlesden-Popper phases La$_{n+1}$Ni$_n$O$_{3n+1}$}},\ }\href {https://doi.org/https://doi.org/10.1016/j.jmst.2023.11.011} {\bibfield  {journal} {\bibinfo  {journal} {Journal of Materials Science \& Technology}\ }\textbf {\bibinfo {volume} {185}},\ \bibinfo {pages} {147} (\bibinfo {year} {2024}{\natexlab{a}})}\BibitemShut {NoStop}%
\bibitem [{\citenamefont {Wang}\ \emph {et~al.}(2023)\citenamefont {Wang}, \citenamefont {Li}, \citenamefont {Xie}, \citenamefont {Liu}, \citenamefont {Sun}, \citenamefont {Huang}, \citenamefont {Gao}, \citenamefont {Nakagawa}, \citenamefont {Fu}, \citenamefont {Dong} \emph {et~al.}}]{wang2023structure}%
  \BibitemOpen
  \bibfield  {author} {\bibinfo {author} {\bibfnamefont {L.}~\bibnamefont {Wang}}, \bibinfo {author} {\bibfnamefont {Y.}~\bibnamefont {Li}}, \bibinfo {author} {\bibfnamefont {S.}~\bibnamefont {Xie}}, \bibinfo {author} {\bibfnamefont {F.}~\bibnamefont {Liu}}, \bibinfo {author} {\bibfnamefont {H.}~\bibnamefont {Sun}}, \bibinfo {author} {\bibfnamefont {C.}~\bibnamefont {Huang}}, \bibinfo {author} {\bibfnamefont {Y.}~\bibnamefont {Gao}}, \bibinfo {author} {\bibfnamefont {T.}~\bibnamefont {Nakagawa}}, \bibinfo {author} {\bibfnamefont {B.}~\bibnamefont {Fu}}, \bibinfo {author} {\bibfnamefont {B.}~\bibnamefont {Dong}}, \emph {et~al.},\ }\bibfield  {title} {\bibinfo {title} {Structure responsible for the superconducting state in {La$_{3}$Ni$_{2}$O$_7$} at low temperature and high pressure conditions},\ }\href@noop {} {\bibfield  {journal} {\bibinfo  {journal} {arXiv preprint arXiv:2311.09186}\ } (\bibinfo {year} {2023})}\BibitemShut {NoStop}%
\bibitem [{\citenamefont {Wang}\ \emph {et~al.}(2024{\natexlab{a}})\citenamefont {Wang}, \citenamefont {Wang}, \citenamefont {Shen}, \citenamefont {Hou}, \citenamefont {Ma}, \citenamefont {Shi}, \citenamefont {Ren}, \citenamefont {Gu}, \citenamefont {Ma}, \citenamefont {Yang}, \citenamefont {Liu}, \citenamefont {Guo}, \citenamefont {Sun}, \citenamefont {Zhang}, \citenamefont {Calder}, \citenamefont {Yan}, \citenamefont {Wang}, \citenamefont {Uwatoko},\ and\ \citenamefont {Cheng}}]{wang2023pressure}%
  \BibitemOpen
  \bibfield  {author} {\bibinfo {author} {\bibfnamefont {G.}~\bibnamefont {Wang}}, \bibinfo {author} {\bibfnamefont {N.~N.}\ \bibnamefont {Wang}}, \bibinfo {author} {\bibfnamefont {X.~L.}\ \bibnamefont {Shen}}, \bibinfo {author} {\bibfnamefont {J.}~\bibnamefont {Hou}}, \bibinfo {author} {\bibfnamefont {L.}~\bibnamefont {Ma}}, \bibinfo {author} {\bibfnamefont {L.~F.}\ \bibnamefont {Shi}}, \bibinfo {author} {\bibfnamefont {Z.~A.}\ \bibnamefont {Ren}}, \bibinfo {author} {\bibfnamefont {Y.~D.}\ \bibnamefont {Gu}}, \bibinfo {author} {\bibfnamefont {H.~M.}\ \bibnamefont {Ma}}, \bibinfo {author} {\bibfnamefont {P.~T.}\ \bibnamefont {Yang}}, \bibinfo {author} {\bibfnamefont {Z.~Y.}\ \bibnamefont {Liu}}, \bibinfo {author} {\bibfnamefont {H.~Z.}\ \bibnamefont {Guo}}, \bibinfo {author} {\bibfnamefont {J.~P.}\ \bibnamefont {Sun}}, \bibinfo {author} {\bibfnamefont {G.~M.}\ \bibnamefont {Zhang}}, \bibinfo {author} {\bibfnamefont {S.}~\bibnamefont {Calder}}, \bibinfo {author} {\bibfnamefont {J.-Q.}\ \bibnamefont {Yan}},
  \bibinfo {author} {\bibfnamefont {B.~S.}\ \bibnamefont {Wang}}, \bibinfo {author} {\bibfnamefont {Y.}~\bibnamefont {Uwatoko}},\ and\ \bibinfo {author} {\bibfnamefont {J.-G.}\ \bibnamefont {Cheng}},\ }\bibfield  {title} {\bibinfo {title} {Pressure-induced superconductivity in polycrystalline {La$_{3}$Ni$_{2}$O$_{7-\delta}$}},\ }\href {https://doi.org/10.1103/PhysRevX.14.011040} {\bibfield  {journal} {\bibinfo  {journal} {Phys. Rev. X}\ }\textbf {\bibinfo {volume} {14}},\ \bibinfo {pages} {011040} (\bibinfo {year} {2024}{\natexlab{a}})}\BibitemShut {NoStop}%
\bibitem [{\citenamefont {Dong}\ \emph {et~al.}(2023)\citenamefont {Dong}, \citenamefont {Huo}, \citenamefont {Li}, \citenamefont {Li}, \citenamefont {Li}, \citenamefont {Sun}, \citenamefont {Lu}, \citenamefont {Wang}, \citenamefont {Wang},\ and\ \citenamefont {Chen}}]{dong2023visualization}%
  \BibitemOpen
  \bibfield  {author} {\bibinfo {author} {\bibfnamefont {Z.}~\bibnamefont {Dong}}, \bibinfo {author} {\bibfnamefont {M.}~\bibnamefont {Huo}}, \bibinfo {author} {\bibfnamefont {J.}~\bibnamefont {Li}}, \bibinfo {author} {\bibfnamefont {J.}~\bibnamefont {Li}}, \bibinfo {author} {\bibfnamefont {P.}~\bibnamefont {Li}}, \bibinfo {author} {\bibfnamefont {H.}~\bibnamefont {Sun}}, \bibinfo {author} {\bibfnamefont {Y.}~\bibnamefont {Lu}}, \bibinfo {author} {\bibfnamefont {M.}~\bibnamefont {Wang}}, \bibinfo {author} {\bibfnamefont {Y.}~\bibnamefont {Wang}},\ and\ \bibinfo {author} {\bibfnamefont {Z.}~\bibnamefont {Chen}},\ }\bibfield  {title} {\bibinfo {title} {Visualization of oxygen vacancies and self-doped ligand holes in {La$_{3}$Ni$_{2}$O$_{7-\delta}$}},\ }\href@noop {} {\bibfield  {journal} {\bibinfo  {journal} {arXiv preprint arXiv:2312.15727}\ } (\bibinfo {year} {2023})}\BibitemShut {NoStop}%
\bibitem [{\citenamefont {Wang}\ \emph {et~al.}(2024{\natexlab{b}})\citenamefont {Wang}, \citenamefont {Wang}, \citenamefont {Shen}, \citenamefont {Hou}, \citenamefont {Luo}, \citenamefont {Ma}, \citenamefont {Yang}, \citenamefont {Shi}, \citenamefont {Dou}, \citenamefont {Feng}, \citenamefont {Yang}, \citenamefont {Shi}, \citenamefont {Ren}, \citenamefont {Ma}, \citenamefont {Yang}, \citenamefont {Liu}, \citenamefont {Liu}, \citenamefont {Zhang}, \citenamefont {Dong},\ and\ \citenamefont {Cheng}}]{WangNature24}%
  \BibitemOpen
  \bibfield  {author} {\bibinfo {author} {\bibfnamefont {N.}~\bibnamefont {Wang}}, \bibinfo {author} {\bibfnamefont {G.}~\bibnamefont {Wang}}, \bibinfo {author} {\bibfnamefont {X.}~\bibnamefont {Shen}}, \bibinfo {author} {\bibfnamefont {J.}~\bibnamefont {Hou}}, \bibinfo {author} {\bibfnamefont {J.}~\bibnamefont {Luo}}, \bibinfo {author} {\bibfnamefont {X.}~\bibnamefont {Ma}}, \bibinfo {author} {\bibfnamefont {H.}~\bibnamefont {Yang}}, \bibinfo {author} {\bibfnamefont {L.}~\bibnamefont {Shi}}, \bibinfo {author} {\bibfnamefont {J.}~\bibnamefont {Dou}}, \bibinfo {author} {\bibfnamefont {J.}~\bibnamefont {Feng}}, \bibinfo {author} {\bibfnamefont {J.}~\bibnamefont {Yang}}, \bibinfo {author} {\bibfnamefont {Y.}~\bibnamefont {Shi}}, \bibinfo {author} {\bibfnamefont {Z.}~\bibnamefont {Ren}}, \bibinfo {author} {\bibfnamefont {H.}~\bibnamefont {Ma}}, \bibinfo {author} {\bibfnamefont {P.}~\bibnamefont {Yang}}, \bibinfo {author} {\bibfnamefont {Z.}~\bibnamefont {Liu}}, \bibinfo {author} {\bibfnamefont {Y.}~\bibnamefont
  {Liu}}, \bibinfo {author} {\bibfnamefont {H.}~\bibnamefont {Zhang}}, \bibinfo {author} {\bibfnamefont {X.}~\bibnamefont {Dong}},\ and\ \bibinfo {author} {\bibfnamefont {J.-G.}\ \bibnamefont {Cheng}},\ }\bibfield  {title} {\bibinfo {title} {Bulk high-temperature superconductivity in pressurized tetragonal {La$_2$PrNi$_2$O$_7$}},\ }\href {https://doi.org/10.1038/s41586-024-07996-8} {\bibfield  {journal} {\bibinfo  {journal} {Nature}\ }\textbf {\bibinfo {volume} {634}},\ \bibinfo {pages} {579} (\bibinfo {year} {2024}{\natexlab{b}})}\BibitemShut {NoStop}%
\bibitem [{\citenamefont {Li}\ \emph {et~al.}(2019)\citenamefont {Li}, \citenamefont {Lee}, \citenamefont {Wang}, \citenamefont {Osada}, \citenamefont {Crossley}, \citenamefont {Lee}, \citenamefont {Cui}, \citenamefont {Hikita},\ and\ \citenamefont {Hwang}}]{Li2019}%
  \BibitemOpen
  \bibfield  {author} {\bibinfo {author} {\bibfnamefont {D.}~\bibnamefont {Li}}, \bibinfo {author} {\bibfnamefont {K.}~\bibnamefont {Lee}}, \bibinfo {author} {\bibfnamefont {B.~Y.}\ \bibnamefont {Wang}}, \bibinfo {author} {\bibfnamefont {M.}~\bibnamefont {Osada}}, \bibinfo {author} {\bibfnamefont {S.}~\bibnamefont {Crossley}}, \bibinfo {author} {\bibfnamefont {H.~R.}\ \bibnamefont {Lee}}, \bibinfo {author} {\bibfnamefont {Y.}~\bibnamefont {Cui}}, \bibinfo {author} {\bibfnamefont {Y.}~\bibnamefont {Hikita}},\ and\ \bibinfo {author} {\bibfnamefont {H.~Y.}\ \bibnamefont {Hwang}},\ }\bibfield  {title} {\bibinfo {title} {Superconductivity in an infinite-layer nickelate},\ }\href {https://doi.org/10.1038/s41586-019-1496-5} {\bibfield  {journal} {\bibinfo  {journal} {Nature}\ }\textbf {\bibinfo {volume} {572}},\ \bibinfo {pages} {624–627} (\bibinfo {year} {2019})}\BibitemShut {NoStop}%
\bibitem [{\citenamefont {Pan}\ \emph {et~al.}(2021)\citenamefont {Pan}, \citenamefont {Ferenc~Segedin}, \citenamefont {LaBollita}, \citenamefont {Song}, \citenamefont {Nica}, \citenamefont {Goodge}, \citenamefont {Pierce}, \citenamefont {Doyle}, \citenamefont {Novakov}, \citenamefont {Córdova~Carrizales}, \citenamefont {N’Diaye}, \citenamefont {Shafer}, \citenamefont {Paik}, \citenamefont {Heron}, \citenamefont {Mason}, \citenamefont {Yacoby}, \citenamefont {Kourkoutis}, \citenamefont {Erten}, \citenamefont {Brooks}, \citenamefont {Botana},\ and\ \citenamefont {Mundy}}]{Pan2021}%
  \BibitemOpen
  \bibfield  {author} {\bibinfo {author} {\bibfnamefont {G.~A.}\ \bibnamefont {Pan}}, \bibinfo {author} {\bibfnamefont {D.}~\bibnamefont {Ferenc~Segedin}}, \bibinfo {author} {\bibfnamefont {H.}~\bibnamefont {LaBollita}}, \bibinfo {author} {\bibfnamefont {Q.}~\bibnamefont {Song}}, \bibinfo {author} {\bibfnamefont {E.~M.}\ \bibnamefont {Nica}}, \bibinfo {author} {\bibfnamefont {B.~H.}\ \bibnamefont {Goodge}}, \bibinfo {author} {\bibfnamefont {A.~T.}\ \bibnamefont {Pierce}}, \bibinfo {author} {\bibfnamefont {S.}~\bibnamefont {Doyle}}, \bibinfo {author} {\bibfnamefont {S.}~\bibnamefont {Novakov}}, \bibinfo {author} {\bibfnamefont {D.}~\bibnamefont {Córdova~Carrizales}}, \bibinfo {author} {\bibfnamefont {A.~T.}\ \bibnamefont {N’Diaye}}, \bibinfo {author} {\bibfnamefont {P.}~\bibnamefont {Shafer}}, \bibinfo {author} {\bibfnamefont {H.}~\bibnamefont {Paik}}, \bibinfo {author} {\bibfnamefont {J.~T.}\ \bibnamefont {Heron}}, \bibinfo {author} {\bibfnamefont {J.~A.}\ \bibnamefont {Mason}}, \bibinfo {author}
  {\bibfnamefont {A.}~\bibnamefont {Yacoby}}, \bibinfo {author} {\bibfnamefont {L.~F.}\ \bibnamefont {Kourkoutis}}, \bibinfo {author} {\bibfnamefont {O.}~\bibnamefont {Erten}}, \bibinfo {author} {\bibfnamefont {C.~M.}\ \bibnamefont {Brooks}}, \bibinfo {author} {\bibfnamefont {A.~S.}\ \bibnamefont {Botana}},\ and\ \bibinfo {author} {\bibfnamefont {J.~A.}\ \bibnamefont {Mundy}},\ }\bibfield  {title} {\bibinfo {title} {Superconductivity in a quintuple-layer square-planar nickelate},\ }\href {https://doi.org/10.1038/s41563-021-01142-9} {\bibfield  {journal} {\bibinfo  {journal} {Nature Materials}\ }\textbf {\bibinfo {volume} {21}},\ \bibinfo {pages} {160–164} (\bibinfo {year} {2021})}\BibitemShut {NoStop}%
\bibitem [{\citenamefont {Osada}\ \emph {et~al.}(2020)\citenamefont {Osada}, \citenamefont {Wang}, \citenamefont {Goodge}, \citenamefont {Lee}, \citenamefont {Yoon}, \citenamefont {Sakuma}, \citenamefont {Li}, \citenamefont {Miura}, \citenamefont {Kourkoutis},\ and\ \citenamefont {Hwang}}]{Osada2020}%
  \BibitemOpen
  \bibfield  {author} {\bibinfo {author} {\bibfnamefont {M.}~\bibnamefont {Osada}}, \bibinfo {author} {\bibfnamefont {B.~Y.}\ \bibnamefont {Wang}}, \bibinfo {author} {\bibfnamefont {B.~H.}\ \bibnamefont {Goodge}}, \bibinfo {author} {\bibfnamefont {K.}~\bibnamefont {Lee}}, \bibinfo {author} {\bibfnamefont {H.}~\bibnamefont {Yoon}}, \bibinfo {author} {\bibfnamefont {K.}~\bibnamefont {Sakuma}}, \bibinfo {author} {\bibfnamefont {D.}~\bibnamefont {Li}}, \bibinfo {author} {\bibfnamefont {M.}~\bibnamefont {Miura}}, \bibinfo {author} {\bibfnamefont {L.~F.}\ \bibnamefont {Kourkoutis}},\ and\ \bibinfo {author} {\bibfnamefont {H.~Y.}\ \bibnamefont {Hwang}},\ }\bibfield  {title} {\bibinfo {title} {A superconducting praseodymium nickelate with infinite layer structure},\ }\href {https://doi.org/10.1021/acs.nanolett.0c01392} {\bibfield  {journal} {\bibinfo  {journal} {Nano Letters}\ }\textbf {\bibinfo {volume} {20}},\ \bibinfo {pages} {5735–5740} (\bibinfo {year} {2020})}\BibitemShut {NoStop}%
\bibitem [{\citenamefont {Lechermann}\ \emph {et~al.}(2023)\citenamefont {Lechermann}, \citenamefont {Gondolf}, \citenamefont {B\"otzel},\ and\ \citenamefont {Eremin}}]{lechermann2023electronic}%
  \BibitemOpen
  \bibfield  {author} {\bibinfo {author} {\bibfnamefont {F.}~\bibnamefont {Lechermann}}, \bibinfo {author} {\bibfnamefont {J.}~\bibnamefont {Gondolf}}, \bibinfo {author} {\bibfnamefont {S.}~\bibnamefont {B\"otzel}},\ and\ \bibinfo {author} {\bibfnamefont {I.~M.}\ \bibnamefont {Eremin}},\ }\bibfield  {title} {\bibinfo {title} {Electronic correlations and superconducting instability in {${\mathrm{La}}_{3}{\mathrm{Ni}}_{2}{\mathrm{O}}_{7}$} under high pressure},\ }\href {https://doi.org/10.1103/PhysRevB.108.L201121} {\bibfield  {journal} {\bibinfo  {journal} {Phys. Rev. B}\ }\textbf {\bibinfo {volume} {108}},\ \bibinfo {pages} {L201121} (\bibinfo {year} {2023})}\BibitemShut {NoStop}%
\bibitem [{\citenamefont {Zhang}\ \emph {et~al.}(2023{\natexlab{b}})\citenamefont {Zhang}, \citenamefont {Lin}, \citenamefont {Moreo},\ and\ \citenamefont {Dagotto}}]{zhang2023electronic}%
  \BibitemOpen
  \bibfield  {author} {\bibinfo {author} {\bibfnamefont {Y.}~\bibnamefont {Zhang}}, \bibinfo {author} {\bibfnamefont {L.-F.}\ \bibnamefont {Lin}}, \bibinfo {author} {\bibfnamefont {A.}~\bibnamefont {Moreo}},\ and\ \bibinfo {author} {\bibfnamefont {E.}~\bibnamefont {Dagotto}},\ }\bibfield  {title} {\bibinfo {title} {Electronic structure, dimer physics, orbital-selective behavior, and magnetic tendencies in the bilayer nickelate superconductor {La$_3$Ni$_2$O$_{7}$} under pressure},\ }\href@noop {} {\bibfield  {journal} {\bibinfo  {journal} {Physical Review B}\ }\textbf {\bibinfo {volume} {108}},\ \bibinfo {pages} {L180510} (\bibinfo {year} {2023}{\natexlab{b}})}\BibitemShut {NoStop}%
\bibitem [{\citenamefont {Liu}\ \emph {et~al.}(2023)\citenamefont {Liu}, \citenamefont {Xia}, \citenamefont {Zhou},\ and\ \citenamefont {Chen}}]{liu2023role}%
  \BibitemOpen
  \bibfield  {author} {\bibinfo {author} {\bibfnamefont {H.}~\bibnamefont {Liu}}, \bibinfo {author} {\bibfnamefont {C.}~\bibnamefont {Xia}}, \bibinfo {author} {\bibfnamefont {S.}~\bibnamefont {Zhou}},\ and\ \bibinfo {author} {\bibfnamefont {H.}~\bibnamefont {Chen}},\ }\bibfield  {title} {\bibinfo {title} {Role of crystal-field-splitting and longe-range-hoppings on superconducting pairing symmetry of {La$_{3}$Ni$_{2}$O$_7$}},\ }\href@noop {} {\bibfield  {journal} {\bibinfo  {journal} {arXiv preprint arXiv:2311.07316}\ } (\bibinfo {year} {2023})}\BibitemShut {NoStop}%
\bibitem [{\citenamefont {Qin}\ and\ \citenamefont {Yang}(2023)}]{qin2023high}%
  \BibitemOpen
  \bibfield  {author} {\bibinfo {author} {\bibfnamefont {Q.}~\bibnamefont {Qin}}\ and\ \bibinfo {author} {\bibfnamefont {Y.-f.}\ \bibnamefont {Yang}},\ }\bibfield  {title} {\bibinfo {title} {High-{T$_c$} superconductivity by mobilizing local spin singlets and possible route to higher {T$_c$} in pressurized {La$_{3}$Ni$_{2}$O$_7$}},\ }\href@noop {} {\bibfield  {journal} {\bibinfo  {journal} {Physical Review B}\ }\textbf {\bibinfo {volume} {108}},\ \bibinfo {pages} {L140504} (\bibinfo {year} {2023})}\BibitemShut {NoStop}%
\bibitem [{\citenamefont {Huang}\ \emph {et~al.}(2023{\natexlab{a}})\citenamefont {Huang}, \citenamefont {Wang},\ and\ \citenamefont {Zhou}}]{huang2023impurity}%
  \BibitemOpen
  \bibfield  {author} {\bibinfo {author} {\bibfnamefont {J.}~\bibnamefont {Huang}}, \bibinfo {author} {\bibfnamefont {Z.}~\bibnamefont {Wang}},\ and\ \bibinfo {author} {\bibfnamefont {T.}~\bibnamefont {Zhou}},\ }\bibfield  {title} {\bibinfo {title} {Impurity and vortex states in the bilayer high-temperature superconductor {La$_{3}$Ni$_{2}$O$_7$}},\ }\href@noop {} {\bibfield  {journal} {\bibinfo  {journal} {Physical Review B}\ }\textbf {\bibinfo {volume} {108}},\ \bibinfo {pages} {174501} (\bibinfo {year} {2023}{\natexlab{a}})}\BibitemShut {NoStop}%
\bibitem [{\citenamefont {Oh}\ and\ \citenamefont {Zhang}(2023)}]{oh2023type}%
  \BibitemOpen
  \bibfield  {author} {\bibinfo {author} {\bibfnamefont {H.}~\bibnamefont {Oh}}\ and\ \bibinfo {author} {\bibfnamefont {Y.-H.}\ \bibnamefont {Zhang}},\ }\bibfield  {title} {\bibinfo {title} {Type-{II $t-J$} model and shared superexchange coupling from hund's rule in superconducting {La$_{3}$Ni$_{2}$O$_7$}},\ }\href {https://doi.org/10.1103/PhysRevB.108.174511} {\bibfield  {journal} {\bibinfo  {journal} {Phys. Rev. B}\ }\textbf {\bibinfo {volume} {108}},\ \bibinfo {pages} {174511} (\bibinfo {year} {2023})}\BibitemShut {NoStop}%
\bibitem [{\citenamefont {Qu}\ \emph {et~al.}(2024)\citenamefont {Qu}, \citenamefont {Qu}, \citenamefont {Chen}, \citenamefont {Wu}, \citenamefont {Yang}, \citenamefont {Li},\ and\ \citenamefont {Su}}]{qu2023bilayer}%
  \BibitemOpen
  \bibfield  {author} {\bibinfo {author} {\bibfnamefont {X.-Z.}\ \bibnamefont {Qu}}, \bibinfo {author} {\bibfnamefont {D.-W.}\ \bibnamefont {Qu}}, \bibinfo {author} {\bibfnamefont {J.}~\bibnamefont {Chen}}, \bibinfo {author} {\bibfnamefont {C.}~\bibnamefont {Wu}}, \bibinfo {author} {\bibfnamefont {F.}~\bibnamefont {Yang}}, \bibinfo {author} {\bibfnamefont {W.}~\bibnamefont {Li}},\ and\ \bibinfo {author} {\bibfnamefont {G.}~\bibnamefont {Su}},\ }\bibfield  {title} {\bibinfo {title} {Bilayer {$t$-$J$-$J_{\perp}$} model and magnetically mediated pairing in the pressurized nickelate {{La$_{3}$Ni$_{2}$O$_{7}$}}},\ }\href@noop {} {\bibfield  {journal} {\bibinfo  {journal} {Physical Review Letters}\ }\textbf {\bibinfo {volume} {132}},\ \bibinfo {pages} {036502} (\bibinfo {year} {2024})}\BibitemShut {NoStop}%
\bibitem [{\citenamefont {Ryee}\ \emph {et~al.}(2023)\citenamefont {Ryee}, \citenamefont {Witt},\ and\ \citenamefont {Wehling}}]{ryee2023critical}%
  \BibitemOpen
  \bibfield  {author} {\bibinfo {author} {\bibfnamefont {S.}~\bibnamefont {Ryee}}, \bibinfo {author} {\bibfnamefont {N.}~\bibnamefont {Witt}},\ and\ \bibinfo {author} {\bibfnamefont {T.~O.}\ \bibnamefont {Wehling}},\ }\bibfield  {title} {\bibinfo {title} {Critical role of interlayer dimer correlations in the superconductivity of {La$_{3}$Ni$_{2}$O$_7$}},\ }\href@noop {} {\bibfield  {journal} {\bibinfo  {journal} {arXiv preprint arXiv:2310.17465}\ } (\bibinfo {year} {2023})}\BibitemShut {NoStop}%
\bibitem [{\citenamefont {Tian}\ \emph {et~al.}(2023)\citenamefont {Tian}, \citenamefont {Chen}, \citenamefont {Wang}, \citenamefont {He},\ and\ \citenamefont {Lu}}]{tian2023correlation}%
  \BibitemOpen
  \bibfield  {author} {\bibinfo {author} {\bibfnamefont {Y.-H.}\ \bibnamefont {Tian}}, \bibinfo {author} {\bibfnamefont {Y.}~\bibnamefont {Chen}}, \bibinfo {author} {\bibfnamefont {J.-M.}\ \bibnamefont {Wang}}, \bibinfo {author} {\bibfnamefont {R.-Q.}\ \bibnamefont {He}},\ and\ \bibinfo {author} {\bibfnamefont {Z.-Y.}\ \bibnamefont {Lu}},\ }\bibfield  {title} {\bibinfo {title} {Correlation effects and concomitant two-orbital $s_{\pm}$-wave superconductivity in la {La$_{3}$Ni$_{2}$O$_7$} under high pressure},\ }\href@noop {} {\bibfield  {journal} {\bibinfo  {journal} {arXiv preprint arXiv:2308.09698}\ } (\bibinfo {year} {2023})}\BibitemShut {NoStop}%
\bibitem [{\citenamefont {Liao}\ \emph {et~al.}(2023)\citenamefont {Liao}, \citenamefont {Chen}, \citenamefont {Duan}, \citenamefont {Wang}, \citenamefont {Liu}, \citenamefont {Yu},\ and\ \citenamefont {Si}}]{liao2023electron}%
  \BibitemOpen
  \bibfield  {author} {\bibinfo {author} {\bibfnamefont {Z.}~\bibnamefont {Liao}}, \bibinfo {author} {\bibfnamefont {L.}~\bibnamefont {Chen}}, \bibinfo {author} {\bibfnamefont {G.}~\bibnamefont {Duan}}, \bibinfo {author} {\bibfnamefont {Y.}~\bibnamefont {Wang}}, \bibinfo {author} {\bibfnamefont {C.}~\bibnamefont {Liu}}, \bibinfo {author} {\bibfnamefont {R.}~\bibnamefont {Yu}},\ and\ \bibinfo {author} {\bibfnamefont {Q.}~\bibnamefont {Si}},\ }\bibfield  {title} {\bibinfo {title} {Electron correlations and superconductivity in {{La$_{3}$Ni$_{2}$O$_{7}$}} under pressure tuning},\ }\href@noop {} {\bibfield  {journal} {\bibinfo  {journal} {arXiv preprint arXiv:2307.16697}\ } (\bibinfo {year} {2023})}\BibitemShut {NoStop}%
\bibitem [{\citenamefont {Kaneko}\ \emph {et~al.}(2024)\citenamefont {Kaneko}, \citenamefont {Sakakibara}, \citenamefont {Ochi},\ and\ \citenamefont {Kuroki}}]{kaneko2023pair}%
  \BibitemOpen
  \bibfield  {author} {\bibinfo {author} {\bibfnamefont {T.}~\bibnamefont {Kaneko}}, \bibinfo {author} {\bibfnamefont {H.}~\bibnamefont {Sakakibara}}, \bibinfo {author} {\bibfnamefont {M.}~\bibnamefont {Ochi}},\ and\ \bibinfo {author} {\bibfnamefont {K.}~\bibnamefont {Kuroki}},\ }\bibfield  {title} {\bibinfo {title} {Pair correlations in the two-orbital hubbard ladder: Implications for superconductivity in the bilayer nickelate {La$_{3}$Ni$_{2}$O$_{7}$}},\ }\href {https://doi.org/10.1103/PhysRevB.109.045154} {\bibfield  {journal} {\bibinfo  {journal} {Phys. Rev. B}\ }\textbf {\bibinfo {volume} {109}},\ \bibinfo {pages} {045154} (\bibinfo {year} {2024})}\BibitemShut {NoStop}%
\bibitem [{\citenamefont {Luo}\ \emph {et~al.}(2023)\citenamefont {Luo}, \citenamefont {Hu}, \citenamefont {Wang}, \citenamefont {W\'u},\ and\ \citenamefont {Yao}}]{LuoModel}%
  \BibitemOpen
  \bibfield  {author} {\bibinfo {author} {\bibfnamefont {Z.}~\bibnamefont {Luo}}, \bibinfo {author} {\bibfnamefont {X.}~\bibnamefont {Hu}}, \bibinfo {author} {\bibfnamefont {M.}~\bibnamefont {Wang}}, \bibinfo {author} {\bibfnamefont {W.}~\bibnamefont {W\'u}},\ and\ \bibinfo {author} {\bibfnamefont {D.-X.}\ \bibnamefont {Yao}},\ }\bibfield  {title} {\bibinfo {title} {Bilayer two-orbital model of {$\mathrm{L}{\mathrm{a}}_{3}\mathrm{N}{\mathrm{i}}_{2}{\mathrm{O}}_{7}$} under pressure},\ }\href {https://doi.org/10.1103/PhysRevLett.131.126001} {\bibfield  {journal} {\bibinfo  {journal} {Phys. Rev. Lett.}\ }\textbf {\bibinfo {volume} {131}},\ \bibinfo {pages} {126001} (\bibinfo {year} {2023})}\BibitemShut {NoStop}%
\bibitem [{\citenamefont {Chen}\ \emph {et~al.}(2023)\citenamefont {Chen}, \citenamefont {Yang},\ and\ \citenamefont {Li}}]{chen2023orbital}%
  \BibitemOpen
  \bibfield  {author} {\bibinfo {author} {\bibfnamefont {J.}~\bibnamefont {Chen}}, \bibinfo {author} {\bibfnamefont {F.}~\bibnamefont {Yang}},\ and\ \bibinfo {author} {\bibfnamefont {W.}~\bibnamefont {Li}},\ }\bibfield  {title} {\bibinfo {title} {Orbital-selective superconductivity in the pressurized bilayer nickelate {{La$_{3}$Ni$_{2}$O$_{7}$}} : An infinite projected entangled-pair state study},\ }\href@noop {} {\bibfield  {journal} {\bibinfo  {journal} {arXiv preprint arXiv:2311.05491}\ } (\bibinfo {year} {2023})}\BibitemShut {NoStop}%
\bibitem [{\citenamefont {Jiang}\ \emph {et~al.}(2023)\citenamefont {Jiang}, \citenamefont {Wang},\ and\ \citenamefont {Zhang}}]{jiang2023high}%
  \BibitemOpen
  \bibfield  {author} {\bibinfo {author} {\bibfnamefont {K.}~\bibnamefont {Jiang}}, \bibinfo {author} {\bibfnamefont {Z.}~\bibnamefont {Wang}},\ and\ \bibinfo {author} {\bibfnamefont {F.-C.}\ \bibnamefont {Zhang}},\ }\bibfield  {title} {\bibinfo {title} {High temperature superconductivity in {La$_{3}$Ni$_{2}$O$_{7}$}},\ }\href@noop {} {\bibfield  {journal} {\bibinfo  {journal} {Chinese Physics Letters}\ } (\bibinfo {year} {2023})}\BibitemShut {NoStop}%
\bibitem [{\citenamefont {Shen}\ \emph {et~al.}(2023)\citenamefont {Shen}, \citenamefont {Qin},\ and\ \citenamefont {Zhang}}]{shen2023effective}%
  \BibitemOpen
  \bibfield  {author} {\bibinfo {author} {\bibfnamefont {Y.}~\bibnamefont {Shen}}, \bibinfo {author} {\bibfnamefont {M.}~\bibnamefont {Qin}},\ and\ \bibinfo {author} {\bibfnamefont {G.-M.}\ \bibnamefont {Zhang}},\ }\bibfield  {title} {\bibinfo {title} {Effective bi-layer model hamiltonian and density-matrix renormalization group study for the high-tc superconductivity in {{La$_{3}$Ni$_{2}$O$_{7}$}} under high pressure},\ }\href {https://doi.org/10.1088/0256-307X/40/12/127401} {\bibfield  {journal} {\bibinfo  {journal} {Chinese Physics Letters}\ }\textbf {\bibinfo {volume} {40}},\ \bibinfo {pages} {127401} (\bibinfo {year} {2023})}\BibitemShut {NoStop}%
\bibitem [{\citenamefont {Yang}\ \emph {et~al.}(2023{\natexlab{a}})\citenamefont {Yang}, \citenamefont {Zhang},\ and\ \citenamefont {Zhang}}]{yang2023minimal}%
  \BibitemOpen
  \bibfield  {author} {\bibinfo {author} {\bibfnamefont {Y.-f.}\ \bibnamefont {Yang}}, \bibinfo {author} {\bibfnamefont {G.-M.}\ \bibnamefont {Zhang}},\ and\ \bibinfo {author} {\bibfnamefont {F.-C.}\ \bibnamefont {Zhang}},\ }\bibfield  {title} {\bibinfo {title} {Interlayer valence bonds and two-component theory for high-${T}_{c}$ superconductivity of {La$_{3}$Ni$_{2}$O$_7$} under pressure},\ }\href {https://doi.org/10.1103/PhysRevB.108.L201108} {\bibfield  {journal} {\bibinfo  {journal} {Phys. Rev. B}\ }\textbf {\bibinfo {volume} {108}},\ \bibinfo {pages} {L201108} (\bibinfo {year} {2023}{\natexlab{a}})}\BibitemShut {NoStop}%
\bibitem [{\citenamefont {W{\'u}}\ \emph {et~al.}(2024)\citenamefont {W{\'u}}, \citenamefont {Luo}, \citenamefont {Yao},\ and\ \citenamefont {Wang}}]{wu2023charge}%
  \BibitemOpen
  \bibfield  {author} {\bibinfo {author} {\bibfnamefont {W.}~\bibnamefont {W{\'u}}}, \bibinfo {author} {\bibfnamefont {Z.}~\bibnamefont {Luo}}, \bibinfo {author} {\bibfnamefont {D.-X.}\ \bibnamefont {Yao}},\ and\ \bibinfo {author} {\bibfnamefont {M.}~\bibnamefont {Wang}},\ }\bibfield  {title} {\bibinfo {title} {Superexchange and charge transfer in the nickelate superconductor $\mathrm{La_3Ni_2O_7}$ under pressure},\ }\href {https://link.springer.com/article/10.1007/s11433-023-2300-4} {\bibfield  {journal} {\bibinfo  {journal} {Science China Physics, Mechanics \& Astronomy}\ }\textbf {\bibinfo {volume} {67}},\ \bibinfo {pages} {117402} (\bibinfo {year} {2024})}\BibitemShut {NoStop}%
\bibitem [{\citenamefont {Luo}\ \emph {et~al.}(2024)\citenamefont {Luo}, \citenamefont {Lv}, \citenamefont {Wang}, \citenamefont {W{\'u}},\ and\ \citenamefont {Yao}}]{luo2024high}%
  \BibitemOpen
  \bibfield  {author} {\bibinfo {author} {\bibfnamefont {Z.}~\bibnamefont {Luo}}, \bibinfo {author} {\bibfnamefont {B.}~\bibnamefont {Lv}}, \bibinfo {author} {\bibfnamefont {M.}~\bibnamefont {Wang}}, \bibinfo {author} {\bibfnamefont {W.}~\bibnamefont {W{\'u}}},\ and\ \bibinfo {author} {\bibfnamefont {D.-X.}\ \bibnamefont {Yao}},\ }\bibfield  {title} {\bibinfo {title} {{High-$T_c$ superconductivity in La$_3$Ni$_2$O$_7$ based on the bilayer two-orbital $tJ$ model}},\ }\href@noop {} {\bibfield  {journal} {\bibinfo  {journal} {npj Quantum Materials}\ }\textbf {\bibinfo {volume} {9}},\ \bibinfo {pages} {61} (\bibinfo {year} {2024})}\BibitemShut {NoStop}%
\bibitem [{\citenamefont {Dagotto}\ \emph {et~al.}(1992)\citenamefont {Dagotto}, \citenamefont {Riera},\ and\ \citenamefont {Scalapino}}]{dagotto1992superconductivity}%
  \BibitemOpen
  \bibfield  {author} {\bibinfo {author} {\bibfnamefont {E.}~\bibnamefont {Dagotto}}, \bibinfo {author} {\bibfnamefont {J.}~\bibnamefont {Riera}},\ and\ \bibinfo {author} {\bibfnamefont {D.}~\bibnamefont {Scalapino}},\ }\bibfield  {title} {\bibinfo {title} {Superconductivity in ladders and coupled planes},\ }\href@noop {} {\bibfield  {journal} {\bibinfo  {journal} {Physical Review B}\ }\textbf {\bibinfo {volume} {45}},\ \bibinfo {pages} {5744} (\bibinfo {year} {1992})}\BibitemShut {NoStop}%
\bibitem [{\citenamefont {Zhang}\ \emph {et~al.}(2023{\natexlab{c}})\citenamefont {Zhang}, \citenamefont {Lin}, \citenamefont {Moreo}, \citenamefont {Maier},\ and\ \citenamefont {Dagotto}}]{zhang2023trends}%
  \BibitemOpen
  \bibfield  {author} {\bibinfo {author} {\bibfnamefont {Y.}~\bibnamefont {Zhang}}, \bibinfo {author} {\bibfnamefont {L.-F.}\ \bibnamefont {Lin}}, \bibinfo {author} {\bibfnamefont {A.}~\bibnamefont {Moreo}}, \bibinfo {author} {\bibfnamefont {T.~A.}\ \bibnamefont {Maier}},\ and\ \bibinfo {author} {\bibfnamefont {E.}~\bibnamefont {Dagotto}},\ }\bibfield  {title} {\bibinfo {title} {Trends in electronic structures and s$\pm$-wave pairing for the rare-earth series in bilayer nickelate superconductor {R$_3$Ni$_2$O$_7$}},\ }\href@noop {} {\bibfield  {journal} {\bibinfo  {journal} {Physical Review B}\ }\textbf {\bibinfo {volume} {108}},\ \bibinfo {pages} {165141} (\bibinfo {year} {2023}{\natexlab{c}})}\BibitemShut {NoStop}%
\bibitem [{\citenamefont {Yang}\ \emph {et~al.}(2023{\natexlab{b}})\citenamefont {Yang}, \citenamefont {Wang},\ and\ \citenamefont {Wang}}]{yang2023possible}%
  \BibitemOpen
  \bibfield  {author} {\bibinfo {author} {\bibfnamefont {Q.-G.}\ \bibnamefont {Yang}}, \bibinfo {author} {\bibfnamefont {D.}~\bibnamefont {Wang}},\ and\ \bibinfo {author} {\bibfnamefont {Q.-H.}\ \bibnamefont {Wang}},\ }\bibfield  {title} {\bibinfo {title} {Possible ${s}_{\ifmmode\pm\else\textpm\fi{}}$-wave superconductivity in {${\mathrm{La}}_{3}{\mathrm{Ni}}_{2}{\mathrm{O}}_{7}$}},\ }\href {https://doi.org/10.1103/PhysRevB.108.L140505} {\bibfield  {journal} {\bibinfo  {journal} {Phys. Rev. B}\ }\textbf {\bibinfo {volume} {108}},\ \bibinfo {pages} {L140505} (\bibinfo {year} {2023}{\natexlab{b}})}\BibitemShut {NoStop}%
\bibitem [{\citenamefont {Zhang}\ \emph {et~al.}(2024{\natexlab{b}})\citenamefont {Zhang}, \citenamefont {Lin}, \citenamefont {Moreo}, \citenamefont {Maier},\ and\ \citenamefont {Dagotto}}]{zhang2023structural}%
  \BibitemOpen
  \bibfield  {author} {\bibinfo {author} {\bibfnamefont {Y.}~\bibnamefont {Zhang}}, \bibinfo {author} {\bibfnamefont {L.-F.}\ \bibnamefont {Lin}}, \bibinfo {author} {\bibfnamefont {A.}~\bibnamefont {Moreo}}, \bibinfo {author} {\bibfnamefont {T.~A.}\ \bibnamefont {Maier}},\ and\ \bibinfo {author} {\bibfnamefont {E.}~\bibnamefont {Dagotto}},\ }\bibfield  {title} {\bibinfo {title} {Structural phase transition, s$\pm$-wave pairing, and magnetic stripe order in bilayered superconductor {La$_3$Ni$_2$O$_{7}$} under pressure},\ }\href {https://doi.org/10.1038/s41467-024-46622-z} {\bibfield  {journal} {\bibinfo  {journal} {Nature Communications}\ }\textbf {\bibinfo {volume} {15}},\ \bibinfo {pages} {2470} (\bibinfo {year} {2024}{\natexlab{b}})}\BibitemShut {NoStop}%
\bibitem [{\citenamefont {Heier}\ \emph {et~al.}(2024)\citenamefont {Heier}, \citenamefont {Park},\ and\ \citenamefont {Savrasov}}]{heier2023competing}%
  \BibitemOpen
  \bibfield  {author} {\bibinfo {author} {\bibfnamefont {G.}~\bibnamefont {Heier}}, \bibinfo {author} {\bibfnamefont {K.}~\bibnamefont {Park}},\ and\ \bibinfo {author} {\bibfnamefont {S.~Y.}\ \bibnamefont {Savrasov}},\ }\bibfield  {title} {\bibinfo {title} {Competing d$_{xy}$ and s$_{\pm }$ pairing symmetries in superconducting {La$_{3}$Ni$_{2}$O$_{7}$}: $\mathrm{LDA}+\mathrm{FLEX}$ calculations},\ }\href {https://doi.org/10.1103/PhysRevB.109.104508} {\bibfield  {journal} {\bibinfo  {journal} {Phys. Rev. B}\ }\textbf {\bibinfo {volume} {109}},\ \bibinfo {pages} {104508} (\bibinfo {year} {2024})}\BibitemShut {NoStop}%
\bibitem [{\citenamefont {Yang}\ \emph {et~al.}(2023{\natexlab{c}})\citenamefont {Yang}, \citenamefont {Oh},\ and\ \citenamefont {Zhang}}]{yang2023strong}%
  \BibitemOpen
  \bibfield  {author} {\bibinfo {author} {\bibfnamefont {H.}~\bibnamefont {Yang}}, \bibinfo {author} {\bibfnamefont {H.}~\bibnamefont {Oh}},\ and\ \bibinfo {author} {\bibfnamefont {Y.-H.}\ \bibnamefont {Zhang}},\ }\bibfield  {title} {\bibinfo {title} {Strong pairing from doping-induced feshbach resonance and second fermi liquid through doping a bilayer spin-one mott insulator: application to {{La$_{3}$Ni$_{2}$O$_{7}$}}},\ }\href@noop {} {\bibfield  {journal} {\bibinfo  {journal} {arXiv preprint arXiv:2309.15095}\ } (\bibinfo {year} {2023}{\natexlab{c}})}\BibitemShut {NoStop}%
\bibitem [{\citenamefont {Sakakibara}\ \emph {et~al.}(2024)\citenamefont {Sakakibara}, \citenamefont {Kitamine}, \citenamefont {Ochi},\ and\ \citenamefont {Kuroki}}]{sakakibara2024possible}%
  \BibitemOpen
  \bibfield  {author} {\bibinfo {author} {\bibfnamefont {H.}~\bibnamefont {Sakakibara}}, \bibinfo {author} {\bibfnamefont {N.}~\bibnamefont {Kitamine}}, \bibinfo {author} {\bibfnamefont {M.}~\bibnamefont {Ochi}},\ and\ \bibinfo {author} {\bibfnamefont {K.}~\bibnamefont {Kuroki}},\ }\bibfield  {title} {\bibinfo {title} {Possible high ${T}_{c}$ superconductivity in {La$_3$Ni$_2$O$_7$} under high pressure through manifestation of a nearly half-filled bilayer hubbard model},\ }\href {https://doi.org/10.1103/PhysRevLett.132.106002} {\bibfield  {journal} {\bibinfo  {journal} {Phys. Rev. Lett.}\ }\textbf {\bibinfo {volume} {132}},\ \bibinfo {pages} {106002} (\bibinfo {year} {2024})}\BibitemShut {NoStop}%
\bibitem [{\citenamefont {Lu}\ \emph {et~al.}(2024)\citenamefont {Lu}, \citenamefont {Pan}, \citenamefont {Yang},\ and\ \citenamefont {Wu}}]{lu2023interlayer}%
  \BibitemOpen
  \bibfield  {author} {\bibinfo {author} {\bibfnamefont {C.}~\bibnamefont {Lu}}, \bibinfo {author} {\bibfnamefont {Z.}~\bibnamefont {Pan}}, \bibinfo {author} {\bibfnamefont {F.}~\bibnamefont {Yang}},\ and\ \bibinfo {author} {\bibfnamefont {C.}~\bibnamefont {Wu}},\ }\bibfield  {title} {\bibinfo {title} {Interlayer-coupling-driven high-temperature superconductivity in {La$_{3}$Ni$_{2}$O$_{7}$} under pressure},\ }\href {https://doi.org/10.1103/PhysRevLett.132.146002} {\bibfield  {journal} {\bibinfo  {journal} {Phys. Rev. Lett.}\ }\textbf {\bibinfo {volume} {132}},\ \bibinfo {pages} {146002} (\bibinfo {year} {2024})}\BibitemShut {NoStop}%
\bibitem [{\citenamefont {Yang}\ \emph {et~al.}(2024)\citenamefont {Yang}, \citenamefont {Sun}, \citenamefont {Hu}, \citenamefont {Xie}, \citenamefont {Miao}, \citenamefont {Luo}, \citenamefont {Chen}, \citenamefont {Liang}, \citenamefont {Zhu}, \citenamefont {Qu} \emph {et~al.}}]{yang2024orbital}%
  \BibitemOpen
  \bibfield  {author} {\bibinfo {author} {\bibfnamefont {J.}~\bibnamefont {Yang}}, \bibinfo {author} {\bibfnamefont {H.}~\bibnamefont {Sun}}, \bibinfo {author} {\bibfnamefont {X.}~\bibnamefont {Hu}}, \bibinfo {author} {\bibfnamefont {Y.}~\bibnamefont {Xie}}, \bibinfo {author} {\bibfnamefont {T.}~\bibnamefont {Miao}}, \bibinfo {author} {\bibfnamefont {H.}~\bibnamefont {Luo}}, \bibinfo {author} {\bibfnamefont {H.}~\bibnamefont {Chen}}, \bibinfo {author} {\bibfnamefont {B.}~\bibnamefont {Liang}}, \bibinfo {author} {\bibfnamefont {W.}~\bibnamefont {Zhu}}, \bibinfo {author} {\bibfnamefont {G.}~\bibnamefont {Qu}}, \emph {et~al.},\ }\bibfield  {title} {\bibinfo {title} {Orbital-dependent electron correlation in double-layer nickelate {La$_3$Ni$_2$O$_7$}},\ }\href@noop {} {\bibfield  {journal} {\bibinfo  {journal} {Nature Communications}\ }\textbf {\bibinfo {volume} {15}},\ \bibinfo {pages} {4373} (\bibinfo {year} {2024})}\BibitemShut {NoStop}%
\bibitem [{\citenamefont {Li}\ \emph {et~al.}(2024)\citenamefont {Li}, \citenamefont {Du}, \citenamefont {Cao}, \citenamefont {Pei}, \citenamefont {Zhang}, \citenamefont {Zhao}, \citenamefont {Zhai}, \citenamefont {Xu}, \citenamefont {Liu}, \citenamefont {Li} \emph {et~al.}}]{li2024electronic}%
  \BibitemOpen
  \bibfield  {author} {\bibinfo {author} {\bibfnamefont {Y.}~\bibnamefont {Li}}, \bibinfo {author} {\bibfnamefont {X.}~\bibnamefont {Du}}, \bibinfo {author} {\bibfnamefont {Y.}~\bibnamefont {Cao}}, \bibinfo {author} {\bibfnamefont {C.}~\bibnamefont {Pei}}, \bibinfo {author} {\bibfnamefont {M.}~\bibnamefont {Zhang}}, \bibinfo {author} {\bibfnamefont {W.}~\bibnamefont {Zhao}}, \bibinfo {author} {\bibfnamefont {K.}~\bibnamefont {Zhai}}, \bibinfo {author} {\bibfnamefont {R.}~\bibnamefont {Xu}}, \bibinfo {author} {\bibfnamefont {Z.}~\bibnamefont {Liu}}, \bibinfo {author} {\bibfnamefont {Z.}~\bibnamefont {Li}}, \emph {et~al.},\ }\bibfield  {title} {\bibinfo {title} {Electronic correlation and pseudogap-like behavior of high-temperature superconductor {La$_3$Ni$_2$O$_7$}},\ }\href@noop {} {\bibfield  {journal} {\bibinfo  {journal} {Chinese Physics Letters}\ }\textbf {\bibinfo {volume} {41}},\ \bibinfo {pages} {087402} (\bibinfo {year} {2024})}\BibitemShut {NoStop}%
\bibitem [{\citenamefont {Zhang}\ \emph {et~al.}(2024{\natexlab{c}})\citenamefont {Zhang}, \citenamefont {Su}, \citenamefont {Huang}, \citenamefont {Shan}, \citenamefont {Sun}, \citenamefont {Huo}, \citenamefont {Ye}, \citenamefont {Zhang}, \citenamefont {Yang}, \citenamefont {Xu} \emph {et~al.}}]{zhang2024high}%
  \BibitemOpen
  \bibfield  {author} {\bibinfo {author} {\bibfnamefont {Y.}~\bibnamefont {Zhang}}, \bibinfo {author} {\bibfnamefont {D.}~\bibnamefont {Su}}, \bibinfo {author} {\bibfnamefont {Y.}~\bibnamefont {Huang}}, \bibinfo {author} {\bibfnamefont {Z.}~\bibnamefont {Shan}}, \bibinfo {author} {\bibfnamefont {H.}~\bibnamefont {Sun}}, \bibinfo {author} {\bibfnamefont {M.}~\bibnamefont {Huo}}, \bibinfo {author} {\bibfnamefont {K.}~\bibnamefont {Ye}}, \bibinfo {author} {\bibfnamefont {J.}~\bibnamefont {Zhang}}, \bibinfo {author} {\bibfnamefont {Z.}~\bibnamefont {Yang}}, \bibinfo {author} {\bibfnamefont {Y.}~\bibnamefont {Xu}}, \emph {et~al.},\ }\bibfield  {title} {\bibinfo {title} {High-temperature superconductivity with zero resistance and strange-metal behaviour in {La$_3$Ni$_2$O$_{7-\delta}$}},\ }\href {https://doi.org/10.1038/s41567-024-02515-y} {\bibfield  {journal} {\bibinfo  {journal} {Nature Physics}\ }\textbf {\bibinfo {volume} {20}},\ \bibinfo {pages} {1269} (\bibinfo {year} {2024}{\natexlab{c}})}\BibitemShut
  {NoStop}%
\bibitem [{\citenamefont {Zhou}\ \emph {et~al.}(2024)\citenamefont {Zhou}, \citenamefont {Guo}, \citenamefont {Cai}, \citenamefont {Sun}, \citenamefont {Wang}, \citenamefont {Zhao}, \citenamefont {Han}, \citenamefont {Chen}, \citenamefont {Chen}, \citenamefont {Wu}, \citenamefont {Ding}, \citenamefont {Xiang}, \citenamefont {kwang Mao},\ and\ \citenamefont {Sun}}]{zhouguo23}%
  \BibitemOpen
  \bibfield  {author} {\bibinfo {author} {\bibfnamefont {Y.}~\bibnamefont {Zhou}}, \bibinfo {author} {\bibfnamefont {J.}~\bibnamefont {Guo}}, \bibinfo {author} {\bibfnamefont {S.}~\bibnamefont {Cai}}, \bibinfo {author} {\bibfnamefont {H.}~\bibnamefont {Sun}}, \bibinfo {author} {\bibfnamefont {P.}~\bibnamefont {Wang}}, \bibinfo {author} {\bibfnamefont {J.}~\bibnamefont {Zhao}}, \bibinfo {author} {\bibfnamefont {J.}~\bibnamefont {Han}}, \bibinfo {author} {\bibfnamefont {X.}~\bibnamefont {Chen}}, \bibinfo {author} {\bibfnamefont {Y.}~\bibnamefont {Chen}}, \bibinfo {author} {\bibfnamefont {Q.}~\bibnamefont {Wu}}, \bibinfo {author} {\bibfnamefont {Y.}~\bibnamefont {Ding}}, \bibinfo {author} {\bibfnamefont {T.}~\bibnamefont {Xiang}}, \bibinfo {author} {\bibfnamefont {H.}~\bibnamefont {kwang Mao}},\ and\ \bibinfo {author} {\bibfnamefont {L.}~\bibnamefont {Sun}},\ }\href {https://arxiv.org/abs/2311.12361} {\bibinfo {title} {{Investigations of key issues on the reproducibility of high-$T_c$ superconductivity emerging
  from compressed La$_3$Ni$_2$O$_7$}}} (\bibinfo {year} {2024}),\ \Eprint {https://arxiv.org/abs/2311.12361} {arXiv:2311.12361 [cond-mat.supr-con]} \BibitemShut {NoStop}%
\bibitem [{\citenamefont {B\"otzel}\ \emph {et~al.}(2024)\citenamefont {B\"otzel}, \citenamefont {Lechermann}, \citenamefont {Gondolf},\ and\ \citenamefont {Eremin}}]{Boetzel2024}%
  \BibitemOpen
  \bibfield  {author} {\bibinfo {author} {\bibfnamefont {S.}~\bibnamefont {B\"otzel}}, \bibinfo {author} {\bibfnamefont {F.}~\bibnamefont {Lechermann}}, \bibinfo {author} {\bibfnamefont {J.}~\bibnamefont {Gondolf}},\ and\ \bibinfo {author} {\bibfnamefont {I.~M.}\ \bibnamefont {Eremin}},\ }\bibfield  {title} {\bibinfo {title} {Theory of magnetic excitations in the multilayer nickelate superconductor {${\mathrm{La}}_{3}{\mathrm{Ni}}_{2}{\mathrm{O}}_{7}$}},\ }\href {https://doi.org/10.1103/PhysRevB.109.L180502} {\bibfield  {journal} {\bibinfo  {journal} {Phys. Rev. B}\ }\textbf {\bibinfo {volume} {109}},\ \bibinfo {pages} {L180502} (\bibinfo {year} {2024})}\BibitemShut {NoStop}%
\bibitem [{\citenamefont {Huang}\ \emph {et~al.}(2023{\natexlab{b}})\citenamefont {Huang}, \citenamefont {Wang},\ and\ \citenamefont {Zhou}}]{HuangPRB23}%
  \BibitemOpen
  \bibfield  {author} {\bibinfo {author} {\bibfnamefont {J.}~\bibnamefont {Huang}}, \bibinfo {author} {\bibfnamefont {Z.~D.}\ \bibnamefont {Wang}},\ and\ \bibinfo {author} {\bibfnamefont {T.}~\bibnamefont {Zhou}},\ }\bibfield  {title} {\bibinfo {title} {Impurity and vortex states in the bilayer high-temperature superconductor {${\mathrm{La}}_{3}{\mathrm{Ni}}_{2}{\mathrm{O}}_{7}$}},\ }\href {https://doi.org/10.1103/PhysRevB.108.174501} {\bibfield  {journal} {\bibinfo  {journal} {Phys. Rev. B}\ }\textbf {\bibinfo {volume} {108}},\ \bibinfo {pages} {174501} (\bibinfo {year} {2023}{\natexlab{b}})}\BibitemShut {NoStop}%
\bibitem [{\citenamefont {Yang}(2024)}]{yang2024possible}%
  \BibitemOpen
  \bibfield  {author} {\bibinfo {author} {\bibfnamefont {Y.-f.}\ \bibnamefont {Yang}},\ }\bibfield  {title} {\bibinfo {title} {{Possible Fano effect and suppression of Andreev reflection in La$_3$Ni$_2$O$_7$}},\ }\href@noop {} {\bibfield  {journal} {\bibinfo  {journal} {arXiv preprint arXiv:2408.14294}\ } (\bibinfo {year} {2024})}\BibitemShut {NoStop}%
\bibitem [{\citenamefont {Efremov}\ \emph {et~al.}(2011)\citenamefont {Efremov}, \citenamefont {Korshunov}, \citenamefont {Dolgov}, \citenamefont {Golubov},\ and\ \citenamefont {Hirschfeld}}]{Efremov2011}%
  \BibitemOpen
  \bibfield  {author} {\bibinfo {author} {\bibfnamefont {D.~V.}\ \bibnamefont {Efremov}}, \bibinfo {author} {\bibfnamefont {M.~M.}\ \bibnamefont {Korshunov}}, \bibinfo {author} {\bibfnamefont {O.~V.}\ \bibnamefont {Dolgov}}, \bibinfo {author} {\bibfnamefont {A.~A.}\ \bibnamefont {Golubov}},\ and\ \bibinfo {author} {\bibfnamefont {P.~J.}\ \bibnamefont {Hirschfeld}},\ }\bibfield  {title} {\bibinfo {title} {Disorder-induced transition between ${s}_{\ifmmode\pm\else\textpm\fi{}}$ and ${s}_{++}$ states in two-band superconductors},\ }\href {https://doi.org/10.1103/PhysRevB.84.180512} {\bibfield  {journal} {\bibinfo  {journal} {Phys. Rev. B}\ }\textbf {\bibinfo {volume} {84}},\ \bibinfo {pages} {180512} (\bibinfo {year} {2011})}\BibitemShut {NoStop}%
\bibitem [{\citenamefont {Hirschfeld}(2015)}]{Hirschfeld2015}%
  \BibitemOpen
  \bibfield  {author} {\bibinfo {author} {\bibfnamefont {P.~J.}\ \bibnamefont {Hirschfeld}},\ }\bibfield  {title} {\bibinfo {title} {{Using gap symmetry and structure to reveal the pairing mechanism in Fe-based superconductors}},\ }\href {https://doi.org/10.1016/j.crhy.2015.10.002} {\bibfield  {journal} {\bibinfo  {journal} {Comptes Rendus. Physique}\ }\textbf {\bibinfo {volume} {17}},\ \bibinfo {pages} {197–231} (\bibinfo {year} {2015})}\BibitemShut {NoStop}%
\bibitem [{\citenamefont {Korshunov}\ \emph {et~al.}(2016)\citenamefont {Korshunov}, \citenamefont {Togushova},\ and\ \citenamefont {Dolgov}}]{Korshunov2016}%
  \BibitemOpen
  \bibfield  {author} {\bibinfo {author} {\bibfnamefont {M.~M.}\ \bibnamefont {Korshunov}}, \bibinfo {author} {\bibfnamefont {Y.~N.}\ \bibnamefont {Togushova}},\ and\ \bibinfo {author} {\bibfnamefont {O.~V.}\ \bibnamefont {Dolgov}},\ }\bibfield  {title} {\bibinfo {title} {Impurities in multiband superconductors},\ }\href {https://doi.org/10.3367/ufne.2016.07.037863} {\bibfield  {journal} {\bibinfo  {journal} {Physics-Uspekhi}\ }\textbf {\bibinfo {volume} {59}},\ \bibinfo {pages} {1211–1240} (\bibinfo {year} {2016})}\BibitemShut {NoStop}%
\bibitem [{\citenamefont {Mizukami}\ \emph {et~al.}(2014)\citenamefont {Mizukami}, \citenamefont {Konczykowski}, \citenamefont {Kawamoto}, \citenamefont {Kurata}, \citenamefont {Kasahara}, \citenamefont {Hashimoto}, \citenamefont {Mishra}, \citenamefont {Kreisel}, \citenamefont {Wang}, \citenamefont {Hirschfeld}, \citenamefont {Matsuda},\ and\ \citenamefont {Shibauchi}}]{Mizukami2014}%
  \BibitemOpen
  \bibfield  {author} {\bibinfo {author} {\bibfnamefont {Y.}~\bibnamefont {Mizukami}}, \bibinfo {author} {\bibfnamefont {M.}~\bibnamefont {Konczykowski}}, \bibinfo {author} {\bibfnamefont {Y.}~\bibnamefont {Kawamoto}}, \bibinfo {author} {\bibfnamefont {S.}~\bibnamefont {Kurata}}, \bibinfo {author} {\bibfnamefont {S.}~\bibnamefont {Kasahara}}, \bibinfo {author} {\bibfnamefont {K.}~\bibnamefont {Hashimoto}}, \bibinfo {author} {\bibfnamefont {V.}~\bibnamefont {Mishra}}, \bibinfo {author} {\bibfnamefont {A.}~\bibnamefont {Kreisel}}, \bibinfo {author} {\bibfnamefont {Y.}~\bibnamefont {Wang}}, \bibinfo {author} {\bibfnamefont {P.~J.}\ \bibnamefont {Hirschfeld}}, \bibinfo {author} {\bibfnamefont {Y.}~\bibnamefont {Matsuda}},\ and\ \bibinfo {author} {\bibfnamefont {T.}~\bibnamefont {Shibauchi}},\ }\bibfield  {title} {\bibinfo {title} {Disorder-induced topological change of the superconducting gap structure in iron pnictides},\ }\href {https://doi.org/10.1038/ncomms6657} {\bibfield  {journal} {\bibinfo  {journal}
  {Nature Communications}\ }\textbf {\bibinfo {volume} {5}},\ \bibinfo {pages} {5657} (\bibinfo {year} {2014})}\BibitemShut {NoStop}%
\bibitem [{\citenamefont {Holb\ae{}k}\ \emph {et~al.}(2023)\citenamefont {Holb\ae{}k}, \citenamefont {Christensen}, \citenamefont {Kreisel},\ and\ \citenamefont {Andersen}}]{HolbaekPRB23}%
  \BibitemOpen
  \bibfield  {author} {\bibinfo {author} {\bibfnamefont {S.~C.}\ \bibnamefont {Holb\ae{}k}}, \bibinfo {author} {\bibfnamefont {M.~H.}\ \bibnamefont {Christensen}}, \bibinfo {author} {\bibfnamefont {A.}~\bibnamefont {Kreisel}},\ and\ \bibinfo {author} {\bibfnamefont {B.~M.}\ \bibnamefont {Andersen}},\ }\bibfield  {title} {\bibinfo {title} {Unconventional superconductivity protected from disorder on the kagome lattice},\ }\href {https://doi.org/10.1103/PhysRevB.108.144508} {\bibfield  {journal} {\bibinfo  {journal} {Phys. Rev. B}\ }\textbf {\bibinfo {volume} {108}},\ \bibinfo {pages} {144508} (\bibinfo {year} {2023})}\BibitemShut {NoStop}%
\bibitem [{\citenamefont {Hofmann}\ \emph {et~al.}(1990)\citenamefont {Hofmann}, \citenamefont {Keller},\ and\ \citenamefont {Kulić}}]{Hofmann1990}%
  \BibitemOpen
  \bibfield  {author} {\bibinfo {author} {\bibfnamefont {U.}~\bibnamefont {Hofmann}}, \bibinfo {author} {\bibfnamefont {J.}~\bibnamefont {Keller}},\ and\ \bibinfo {author} {\bibfnamefont {M.}~\bibnamefont {Kulić}},\ }\bibfield  {title} {\bibinfo {title} {Interlayer pairing in high temperature superconductors: effect of nonmagnetic impurities},\ }\href {https://doi.org/10.1007/bf01454209} {\bibfield  {journal} {\bibinfo  {journal} {Zeitschrift f\"{u}r Physik B Condensed Matter}\ }\textbf {\bibinfo {volume} {81}},\ \bibinfo {pages} {25–32} (\bibinfo {year} {1990})}\BibitemShut {NoStop}%
\bibitem [{\citenamefont {Abrikosov}\ and\ \citenamefont {Gor’Kov}(1961)}]{abrikosov1961zh}%
  \BibitemOpen
  \bibfield  {author} {\bibinfo {author} {\bibfnamefont {A.}~\bibnamefont {Abrikosov}}\ and\ \bibinfo {author} {\bibfnamefont {L.}~\bibnamefont {Gor’Kov}},\ }\bibfield  {title} {\bibinfo {title} {Zh. {\'e} ksp. teor. fiz. 39, 1781 1960 sov. phys},\ }\href@noop {} {\bibfield  {journal} {\bibinfo  {journal} {JETP}\ }\textbf {\bibinfo {volume} {12}},\ \bibinfo {pages} {1243} (\bibinfo {year} {1961})}\BibitemShut {NoStop}%
\bibitem [{\citenamefont {Maier}\ and\ \citenamefont {Scalapino}(2011)}]{Maier2011}%
  \BibitemOpen
  \bibfield  {author} {\bibinfo {author} {\bibfnamefont {T.~A.}\ \bibnamefont {Maier}}\ and\ \bibinfo {author} {\bibfnamefont {D.~J.}\ \bibnamefont {Scalapino}},\ }\bibfield  {title} {\bibinfo {title} {Pair structure and the pairing interaction in a bilayer hubbard model for unconventional superconductivity},\ }\bibfield  {journal} {\bibinfo  {journal} {Physical Review B}\ }\textbf {\bibinfo {volume} {84}},\ \href {https://doi.org/10.1103/physrevb.84.180513} {10.1103/physrevb.84.180513} (\bibinfo {year} {2011})\BibitemShut {NoStop}%
\bibitem [{Sup()}]{Supplement}%
  \BibitemOpen
  \href@noop {} {\bibinfo {title} {See {Supplemental Material} for additional material on the mixed interlayer and intralayer $s$-wave, the role and symmetry of the interorbital gaps, an altered single orbital model and additional information on the numeric calculations.}}\BibitemShut {Stop}%
\bibitem [{\citenamefont {Chen}\ \emph {et~al.}(2024)\citenamefont {Chen}, \citenamefont {Liu}, \citenamefont {Jiao}, \citenamefont {Zou}, \citenamefont {Jiang}, \citenamefont {Li}, \citenamefont {Luo}, \citenamefont {Wu}, \citenamefont {Zhang}, \citenamefont {Guo},\ and\ \citenamefont {Shu}}]{327sdw2024}%
  \BibitemOpen
  \bibfield  {author} {\bibinfo {author} {\bibfnamefont {K.}~\bibnamefont {Chen}}, \bibinfo {author} {\bibfnamefont {X.}~\bibnamefont {Liu}}, \bibinfo {author} {\bibfnamefont {J.}~\bibnamefont {Jiao}}, \bibinfo {author} {\bibfnamefont {M.}~\bibnamefont {Zou}}, \bibinfo {author} {\bibfnamefont {C.}~\bibnamefont {Jiang}}, \bibinfo {author} {\bibfnamefont {X.}~\bibnamefont {Li}}, \bibinfo {author} {\bibfnamefont {Y.}~\bibnamefont {Luo}}, \bibinfo {author} {\bibfnamefont {Q.}~\bibnamefont {Wu}}, \bibinfo {author} {\bibfnamefont {N.}~\bibnamefont {Zhang}}, \bibinfo {author} {\bibfnamefont {Y.}~\bibnamefont {Guo}},\ and\ \bibinfo {author} {\bibfnamefont {L.}~\bibnamefont {Shu}},\ }\bibfield  {title} {\bibinfo {title} {Evidence of spin density waves in {${\mathrm{La}}_{3}{\mathrm{Ni}}_{2}{\mathrm{O}}_{7\ensuremath{-}\ensuremath{\delta}}$}},\ }\href {https://doi.org/10.1103/PhysRevLett.132.256503} {\bibfield  {journal} {\bibinfo  {journal} {Phys. Rev. Lett.}\ }\textbf {\bibinfo {volume} {132}},\ \bibinfo {pages}
  {256503} (\bibinfo {year} {2024})}\BibitemShut {NoStop}%
\bibitem [{\citenamefont {Lechermann}\ \emph {et~al.}(2024)\citenamefont {Lechermann}, \citenamefont {B\"otzel},\ and\ \citenamefont {Eremin}}]{lechermann24}%
  \BibitemOpen
  \bibfield  {author} {\bibinfo {author} {\bibfnamefont {F.}~\bibnamefont {Lechermann}}, \bibinfo {author} {\bibfnamefont {S.}~\bibnamefont {B\"otzel}},\ and\ \bibinfo {author} {\bibfnamefont {I.~M.}\ \bibnamefont {Eremin}},\ }\bibfield  {title} {\bibinfo {title} {{Electronic instability, layer selectivity, and Fermi arcs in ${\text{La}}_{3}{\text{Ni}}_{2}{\text{O}}_{7}$}},\ }\href {https://doi.org/10.1103/PhysRevMaterials.8.074802} {\bibfield  {journal} {\bibinfo  {journal} {Phys. Rev. Mater.}\ }\textbf {\bibinfo {volume} {8}},\ \bibinfo {pages} {074802} (\bibinfo {year} {2024})}\BibitemShut {NoStop}%
\bibitem [{\citenamefont {Wang}\ \emph {et~al.}(2024{\natexlab{c}})\citenamefont {Wang}, \citenamefont {Jiang}, \citenamefont {Wang}, \citenamefont {Zhang},\ and\ \citenamefont {Hu}}]{wang2024electronicmagneticstructuresbilayer}%
  \BibitemOpen
  \bibfield  {author} {\bibinfo {author} {\bibfnamefont {Y.}~\bibnamefont {Wang}}, \bibinfo {author} {\bibfnamefont {K.}~\bibnamefont {Jiang}}, \bibinfo {author} {\bibfnamefont {Z.}~\bibnamefont {Wang}}, \bibinfo {author} {\bibfnamefont {F.-C.}\ \bibnamefont {Zhang}},\ and\ \bibinfo {author} {\bibfnamefont {J.}~\bibnamefont {Hu}},\ }\href {https://arxiv.org/abs/2401.15097} {\bibinfo {title} {The electronic and magnetic structures of bilayer {La$_3$Ni$_2$O$_7$} at ambient pressure}} (\bibinfo {year} {2024}{\natexlab{c}}),\ \Eprint {https://arxiv.org/abs/2401.15097} {arXiv:2401.15097 [cond-mat.supr-con]} \BibitemShut {NoStop}%
\bibitem [{\citenamefont {Kogan}\ and\ \citenamefont {Prozorov}(2023)}]{Kogan2023}%
  \BibitemOpen
  \bibfield  {author} {\bibinfo {author} {\bibfnamefont {V.~G.}\ \bibnamefont {Kogan}}\ and\ \bibinfo {author} {\bibfnamefont {R.}~\bibnamefont {Prozorov}},\ }\bibfield  {title} {\bibinfo {title} {Disorder-dependent slopes of the upper critical field in nodal and nodeless superconductors},\ }\href {https://doi.org/10.1103/PhysRevB.108.064502} {\bibfield  {journal} {\bibinfo  {journal} {Phys. Rev. B}\ }\textbf {\bibinfo {volume} {108}},\ \bibinfo {pages} {064502} (\bibinfo {year} {2023})}\BibitemShut {NoStop}%
\end{thebibliography}%


\begin{thebibliography}{1}%
\makeatletter
\providecommand \@ifxundefined [1]{%
 \@ifx{#1\undefined}
}%
\providecommand \@ifnum [1]{%
 \ifnum #1\expandafter \@firstoftwo
 \else \expandafter \@secondoftwo
 \fi
}%
\providecommand \@ifx [1]{%
 \ifx #1\expandafter \@firstoftwo
 \else \expandafter \@secondoftwo
 \fi
}%
\providecommand \natexlab [1]{#1}%
\providecommand \enquote  [1]{``#1''}%
\providecommand \bibnamefont  [1]{#1}%
\providecommand \bibfnamefont [1]{#1}%
\providecommand \citenamefont [1]{#1}%
\providecommand \href@noop [0]{\@secondoftwo}%
\providecommand \href [0]{\begingroup \@sanitize@url \@href}%
\providecommand \@href[1]{\@@startlink{#1}\@@href}%
\providecommand \@@href[1]{\endgroup#1\@@endlink}%
\providecommand \@sanitize@url [0]{\catcode `\\12\catcode `\$12\catcode `\&12\catcode `\#12\catcode `\^12\catcode `\_12\catcode `\%12\relax}%
\providecommand \@@startlink[1]{}%
\providecommand \@@endlink[0]{}%
\providecommand \url  [0]{\begingroup\@sanitize@url \@url }%
\providecommand \@url [1]{\endgroup\@href {#1}{\urlprefix }}%
\providecommand \urlprefix  [0]{URL }%
\providecommand \Eprint [0]{\href }%
\providecommand \doibase [0]{https://doi.org/}%
\providecommand \selectlanguage [0]{\@gobble}%
\providecommand \bibinfo  [0]{\@secondoftwo}%
\providecommand \bibfield  [0]{\@secondoftwo}%
\providecommand \translation [1]{[#1]}%
\providecommand \BibitemOpen [0]{}%
\providecommand \bibitemStop [0]{}%
\providecommand \bibitemNoStop [0]{.\EOS\space}%
\providecommand \EOS [0]{\spacefactor3000\relax}%
\providecommand \BibitemShut  [1]{\csname bibitem#1\endcsname}%
\let\auto@bib@innerbib\@empty
%</preamble>
\bibitem [{\citenamefont {Korshunov}\ \emph {et~al.}(2016)\citenamefont {Korshunov}, \citenamefont {Togushova},\ and\ \citenamefont {Dolgov}}]{Korshunov2016}%
  \BibitemOpen
  \bibfield  {author} {\bibinfo {author} {\bibfnamefont {M.~M.}\ \bibnamefont {Korshunov}}, \bibinfo {author} {\bibfnamefont {Y.~N.}\ \bibnamefont {Togushova}},\ and\ \bibinfo {author} {\bibfnamefont {O.~V.}\ \bibnamefont {Dolgov}},\ }\bibfield  {title} {\bibinfo {title} {Impurities in multiband superconductors},\ }\href {https://doi.org/10.3367/ufne.2016.07.037863} {\bibfield  {journal} {\bibinfo  {journal} {Physics-Uspekhi}\ }\textbf {\bibinfo {volume} {59}},\ \bibinfo {pages} {1211–1240} (\bibinfo {year} {2016})}\BibitemShut {NoStop}%
\end{thebibliography}%
\end{document}

%\pagenumbering{Alph}

\title{Supplemental Material: Theory of potential impurity scattering in pressurized superconducting La\textsubscript{3}Ni\textsubscript{2}O\textsubscript{7}}
\author{Steffen B\"otzel}
\affiliation{Institut f\"ur Theoretische Physik III, Ruhr-Universit\"at Bochum,
  D-44780 Bochum, Germany}
\author{Frank Lechermann}
\affiliation{Institut f\"ur Theoretische Physik III, Ruhr-Universit\"at Bochum,
  D-44780 Bochum, Germany}

\author{Takasada Shibauchi}
\affiliation{Department of Advanced Materials Science, The University of Tokyo, Kashiwa, Chiba 277-8561, Japan}
  
\author{Ilya M. Eremin}
\affiliation{Institut f\"ur Theoretische Physik III, Ruhr-Universit\"at Bochum,
  D-44780 Bochum, Germany}

%\linenumbers  %adds lines to reference changes
\maketitle

\subsection{Mixed solution and role of interorbital components for multiorbital La-327 model}
In this section, additional information is given for the La-327 multiorbital model. We first consider the mixing of the interlayer gap component $\Delta_{\perp}$  with the intralayer component, {\it i.e.} extended $s$-wave component, $\Delta_s$. In addition, we investigate the role of the interorbital components to the gap function and show how the gap function is renormalized in the disordered state in the process of the crossover from the $s_\pm$-wave to $s_{++}$-wave states, respectively.

\begin{figure*}[]
      \includegraphics[width=\linewidth]{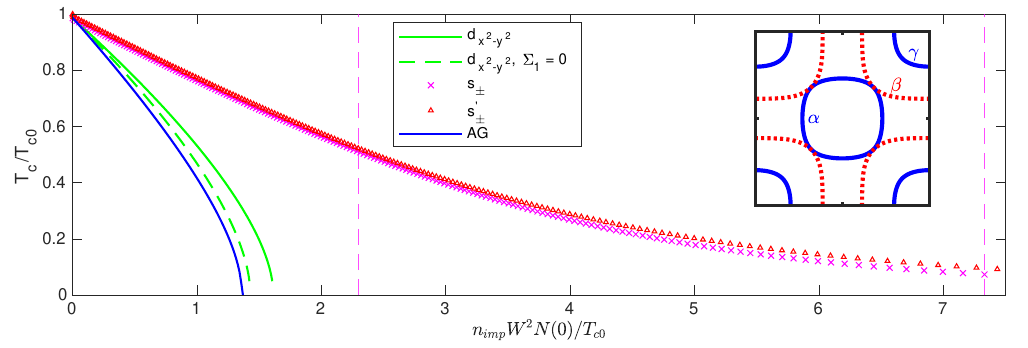}
	\caption{(color online) Calculated $T_c/T_{c0}$ suppression for the two-orbital bilayer La$_3$Ni$_2$O$_7$ model  for the full $s_\pm'$ wave solution $\Delta_{b/a} = \Delta_s \gamma_s \pm \Delta_\perp $ as compared to the pure interlayer $s_\pm$ solution  $\Delta_{b/a} = \pm \Delta_\perp$. We also plot the results for the in-plane $d_{x^2-y^2}$-wave state, where the anomalous self energy is set to zero for simplicity. An Abrikosov-Gor'kov (AG) curve is shown for comparison. The vertical dashed lines denote the impurity densities for which the anomalous self-energy is plotted in Fig.~\ref{fig2SM}(d-f) and Fig.~\ref{fig2SM}(g-i).} \label{fig1SM}
\end{figure*}

In particular, in Fig.~\ref{fig1SM} we compare the effect of non-magnetic impurities for the simple bonding-antibonding~{(ba-)$s_\pm$-wave} (magenta crosses) and the full $s_\pm'$ wave solution $\Delta_{b/a} = \Delta_s \gamma_s \pm \Delta_\perp $ (red triangles) with $\gamma_{s/d} = (\cos(k_x)\pm\cos(k_y))/2$. The intralayer component $\Delta_s$ remains consistently significantly lower than $\Delta_\perp$ across all impurity densities considered, resulting in minimal impact on the shape of the $T_c$ suppression curve. However, it is important to observe that the intralayer component exhibits a tendency to further mitigate the suppression of the unconventional $s$-wave superconductivity with increasing impurity density.

%%%%%%%%%%%%%%%%%%%%%%%%%%%%%%%%%%%%%%%%%%%%%%%%%%%%%%%%%%%%%
%%%%%%%%%%%%%%%%%%%%%%%%%%%%%%%%%%%%%%%%%%%%%%%%%%%%%%%%%%%%%%

\begin{figure*}[]
      \includegraphics[width=\linewidth]{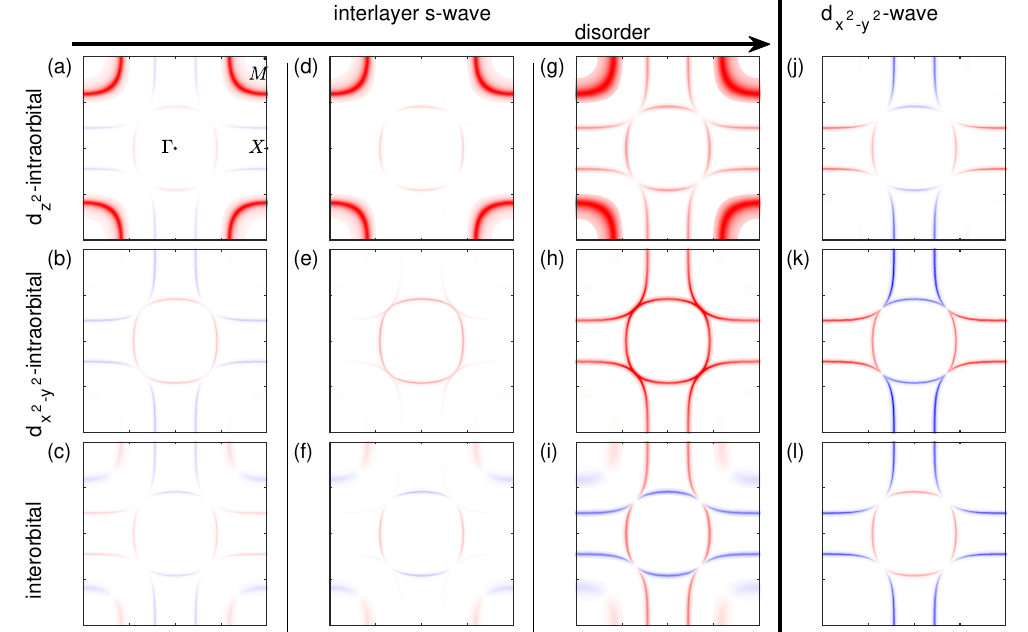}
	\caption{(color online) Different orbital components of the anomalous Green's function in ba-space near $T_c$ for various disorder levels. The first column (a)-(c) correspond to the ba-$s_\pm$-wave in the clean case. The second column (d)-(f) correspond to the ba-$s$-wave in the disordered state, as shown by the vertical dashed line in Fig.S1. The third column (g)-(i) visualizes the ba-$s_{++}$-wave after the $s_{\pm}$ to $s_{++}$ crossover already took place. The corresponding impurity densities for (d)-(f) and (g)-(i) are indicated by the vertical magneta dashed lines in Fig.~\ref{fig1SM}). The fourth column shows the $d_{x^2-y^2}$ solution for the clean case, but the the structure does not qualitatively change upon adding disorder. Red and blue color scheme visualizes the sign and size of the superconducting gap function. The color scheme is fixed for each column but is different between different columns.} \label{fig2SM}
\end{figure*}

To elucidate the role of the hybridization between the $d_{x^2-y^2}$ and $d_{z^2}$ orbitals we note that the $d_{x^2-y^2}$-wave symmetric superconducting gap function with intraorbital gap $\Delta^{b/a}_{x^2-y^2}$ induces an interorbital gap $\Delta^{b/a}_{\text{inter}}$ of the form $\gamma_d^2$. This can be seen by looking on the analytic form of the Green's function given in the Eq.~(8) of the main text. Neglecting the impurity part, it reads 
\begin{equation}
    \hat{G}_{b/a}^{-1}(\mathbf{k},i\omega_n) = i\omega_n \hat{\mathbb{1}} -\hat{H}^{b/a}(\mathbf{k})\hat{\tau}_3 - \hat{\Delta}^{b/a}(\mathbf{k})\hat{\tau}_1,
    \label{Eq:baGFclean}
\end{equation}
where $\hat{\tau}_i$ denotes the $i-$th Pauli matrix in the Gor'kov-Nambu space. Starting with only $\Delta^{b/a}_{x^2-y^2} \neq 0$ and using the  Cramer's rule, the interorbital anomalous component  is given by 
\begin{equation}
    F^{b/a}_{\text{inter}}(\mathbf{k},i\omega_n) = -V_\mathbf{k}^{b/a} \Delta^{b/a}_{x^2-y^2} (\epsilon^{b/a}_{z^2,\mathbf{k}}+ i\omega_n)/\det(\hat{G}_{b/a}^{-1}(\mathbf{k},i\omega_n)),
\end{equation}
where $V_\mathbf{k}^{b/a}$ denotes the hybridization term and $\epsilon^{b/a}_{z^2,\mathbf{k}}$ denotes the intraorbital $d_{z^2}$ component of $\hat{H}^{b/a}(\mathbf{k})$. Both, $V_\mathbf{k}^{b/a}$ and $\Delta^{b/a}_{x^2-y^2}$ contribute a form factor $\gamma_d$. Further, for dominant interlayer hopping, $\epsilon^{b}_{z^2,\mathbf{k}}$ and $\epsilon^{a}_{z^2,\mathbf{k}}$ will be of different sign. Starting with only constant $\Delta^{b/a}_{z^2} \neq 0$, one similarly finds that $\Delta^{b/a}_{\text{inter}}$ follows $\gamma_d$ due to hybridization and the overall $s-$wave symmetry is restored as the total contribution to the gap is proportional to $\gamma^2_d$. We plot the real part of the different orbital components of the anomalous Green's function in ba-space for the lowest Matsubara frequency for the $d_{x^2-y^2}$-wave and also the ba-$s_\pm$-wave in Fig.~\ref{fig2SM}. This quantity is directly related to the superconducting gap function and importantly has the same symmetry in momentum space. Since we calculated near $T_c$, absolute values are small and only the relative values within each column (color scheme is only fixed within each column) are meaningful. In Fig.~\ref{fig2SM}(a-c) and Fig.~\ref{fig2SM}(j-l), the different orbital components are shown for the clean state for the ba-$s_\pm$ and $d$-wave solutions. The interorbital components shown in the lowest row have a different symmetry as the intraorbital components as illustrated by the calculation above. For the ba-$s_\pm$ the largest gap value is realized on the $\gamma$ pocket, whereas it does not participate in the $d$-wave solution. This is expected because the $\gamma$ pocket is of bonding-$d_{z^2}$ character and the $d$-wave solution is in in-plane $d_{x^2-y^2}$-orbital channel and the ba-$s_\pm$-wave is in the interlayer $d_{z^2}$-orbital channel. This also explains the absence of the intraorbital $d_{x^2-y^2}$ component on the $\gamma$ pocket displayed in the second row.
%
%%%%%%%%%%%%%%%%%%%%%%%%%%%%%%%%%%%%%%%%%%%%%%%%%%%%%%%%%%%%%
%%%%%%%%%%%%%%%%%%%%%%%%%%%%%%%%%%%%%%%%%%%%%%%%%%%%%%%%%%%%%%

We shall now incorporate disorder into the analysis. In Fig.~\ref{fig1SM}, we compare $T_c$ suppression curves for the $d_{x^2-y^2}$-wave with and without neglecting the non-vanishing anomalous self-energy due to the interorbital component. The anomalous self-energy decreases the suppression rate slightly but without any qualitative change compared to Abrikosov-Gor'kov (AG) pair-breaking behavior. Since the interorbital component changes sign between bonding $\alpha$ and antibonding $\beta$ band (Fig.~\ref{fig2SM}(l)), this is exactly what is naively expected from the discussion of the ba-$s$-wave. However, it is in principle an interesting question whether this could be important in other systems. Although the suppression rate slightly decreases, the relative values of the different orbital components of superconducting gap do not significantly change compared to the clean case, presented in the fourth column in Fig.~\ref{fig2SM}(j)-(l). As already stated in the main text, the ba-$s$-wave solution displays a $s_\pm\rightarrow s_{++}$ crossover for the intraorbital components. This can be seen in the second and third columns of Fig.~\ref{fig2SM}. The superconducting gap function on the antibonding band is suppressed more strongly than on the dominant bonding bands and vanishes at some points as can be seen in the second row Fig.~\ref{fig2SM}(d)-(f). For higher disorder concentration, the superconducting gap function changes to $s_{++}$ gap structure with the sign dictated by the dominant bonding subspace. Note that the amplitude of the gap function becomes more similar on the different orbital components and on different bands.
 
\subsection{Mixed solution for multiorbital La-327 model and absence of $\gamma$ pocket}
\begin{figure}[]
      \includegraphics[width=\linewidth]{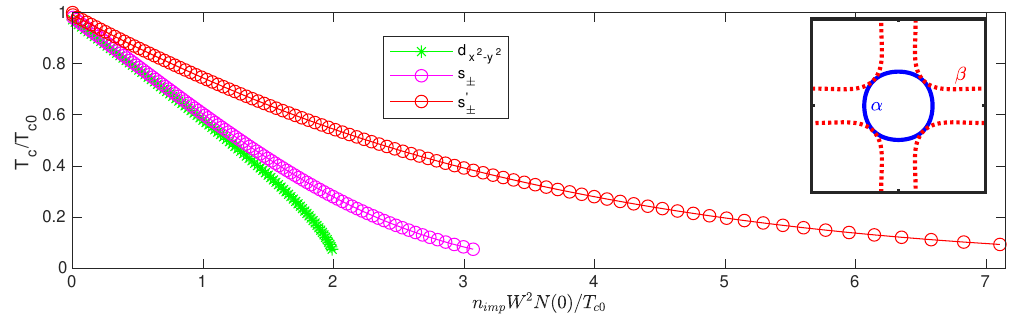}
	\caption{(color online) $T_c$ suppression for two-orbital bilayer La$_3$Ni$_2$O$_7$ model as in Fig.2(b) of the main text. Here, the full $s_\pm'$ wave solution $\Delta_{b/a} = \Delta_s \gamma_s \pm \Delta_\perp $ is shown for comparison with the pure interlayer $s_\pm$ solution  $\Delta_{b/a} = \pm \Delta_\perp$. Here, the intra-layer component of the $s_\pm'$ solution is dominant by an order of magnitude.} \label{fig3SM}
\end{figure}

In this section, the mixing of the interlayer gap component $\Delta_{\perp}$ with the extended intralayer component $\Delta_s$ is considered for the adapted multiorbital La-327 model, in which the $\gamma$ band is pushed 50 meV below the Fermi level. In Fig.~\ref{fig3SM} the $T_c$ suppression for two-orbital bilayer La$_3$Ni$_2$O$_7$ model from the main text (Fig.~2(b)) is compared to the mixed $s'$ solution with $\Delta_{b/a} = \Delta_s \gamma_s \pm \Delta_\perp $ (red circles). For the mixed solution, we fix $J_\perp/J \approx 1.73$ to realistic values and adopt the interaction such that $T_c \approx 80$ K. In contrast to the case where the $\gamma$ pocket is present on the Fermi surface, the intralayer $\Delta_s$ component becomes dominant, mediating a significant decrease in the suppression rate.

For comparison, we also investigated how the pure intralayer $s$-wave reacts to disorder in a single orbital case. By introducing $\Omega_{b/a}^{'}$ implicitly via
\begin{equation}
    \Sigma_1(\omega_n) = \frac{1}{2} n_{\text{imp}} W^{2}  \left[ \tilde{\Delta}_{b} \sum_{\mathbf{k}}  \frac{\gamma_s}{\tilde{\omega}_{n}^{2}+\tilde{\epsilon}_{\mathbf{k},b}^{2}} +  (b \leftrightarrow a)
    \right]   =: \tilde{\Delta}_{b} \Omega_{b}^{'} + \tilde{\Delta}_{a} \Omega^{'}_{a},
    \label{Eq:selfEnergyIntraSTcLimit}
\end{equation}
and we find the analogue to Eq.~(18) of the main text. Inserting now $\Delta_{b/a} = \Delta_s \gamma_s$, the solution for $\Tilde{\Delta}_{b/a}$ becomes
\begin{equation}
    \Tilde{\Delta}_{b/a} = \frac{\Delta_s \gamma_s}{1-\Omega_b^{'} - \Omega_a^{'}}.
\end{equation}
Without the $\gamma_s$ form factor in the definition of $\Omega_{b/a}^{'}$, this would together with  $\tilde{\omega}_{n} = \omega_n/(1-\Omega_{b}-\Omega_{a})$ correspond to the expression from which the Anderson theorem can be derived \cite{Korshunov2016}. However, due to the $\gamma_s$ form factor the suppression rate will depend on the details of the band structure.

%%%%%%%%%%%%%%%%%%%%%%%%%%%%%%%%%%%%%%%%%%%%%%%%%%%%%%%%%%%%%
%%%%%%%%%%%%%%%%%%%%%%%%%%%%%%%%%%%%%%%%%%%%%%%%%%%%%%%%%%%%%%

\subsection{Alternative single orbital model and fixed $t_\perp$}
In this section, we repeat the calculation for another version of the single orbital model. To be precise, a momentum dependence to the interlayer hopping $t_\perp = t_{\perp,0} (\cos(k_x)-\cos(ky))^2$ is added. The resulting band structure loosely resembles the $\alpha$ and $\beta$ pockets for La-327 near quarter filling. Moreover, we maintain $t_{\perp,0} = t$ constant while exclusively varying $J$ and $J_\perp$, in order to achieve an approximately equivalent $T_c$ within the clean case. As in the main text, we consider half-filling $n = 1$, $n = 0.75$ and quarter filling $n=0.5$. The corresponding $T_c$ suppression curves are shown in  Fig.~\ref{fig3SM}. For the given momentum dependence $\epsilon^{b}_{\mathbf{k}} = -\epsilon^{a}_{\mathbf{k}+(\pi,\pi)}$ still holds at half-filling, which again ensures $\Omega_{b} = \Omega_{a}$ and consequently AG behavior at half-filling. The evolution of increasing imbalance between $\Omega_{b}$ and $\Omega_{a}$ decreasing the suppression rate is similar to the case considered in the main text. 
%%%%%%%%%%%%%%%%%%%%%%%%%%%%%%%%%%%%%%%%%%%%%%%%%%%%%%%%%%%%%
\begin{figure*}[]
      \includegraphics[width=\linewidth]{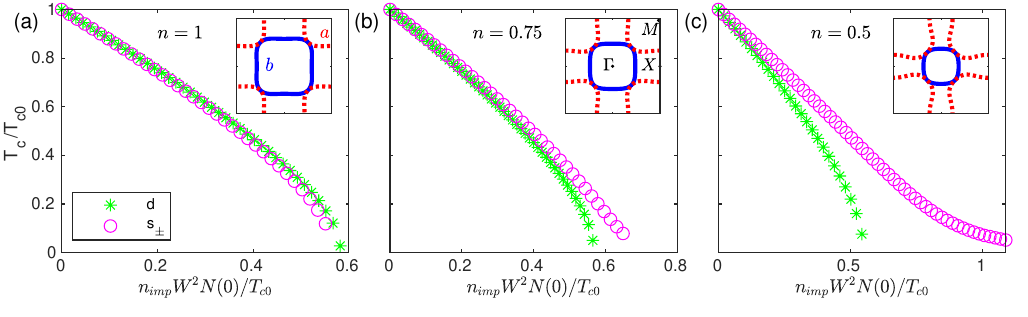}
	\caption{(color online) $T_c$ suppression for a single orbital bilayer model for $\Delta_{b/a} = \pm \Delta_\perp$ and $\Delta_{b/a} = \Delta_d \gamma_d$ for different fillings similar as for Fig.1 of the main text. Here, $t_\perp$ is fixed and $J$ and $J_\perp$ are varied such that $T_c$ is roughly the same for all cases.} \label{fig4SM}
\end{figure*}	
%%%%%%%%%%%%%%%%%%%%%%%%%%%%%%%%%%%%%%%%%%%%%%%%%%%%%%%%%%%%%%
%
Let us for completeness mention, that the pure intralayer $s$-wave described in the previous section also becomes dominant at quarter filling (while vanishing for the other considered fillings) for both variations of the single orbital model. 

\subsection{Computational details}
\label{Sec:CompDet}
The Matsubara frequency cut-off for the summation over Matsubara frequencies is chosen as 16 eV. The quasiparticle energies and gap functions are discretized on a 100$\times$100 k-mesh.

For the single orbital model, we self-consistently solve the linearized versions of Eq.(11)/(12) and Eqs.(14-16) of the main text and restrict ourselves to the pure interlayer ba-$s_\pm$-wave and $d$-wave case. In Fig.~1, we employ $t=0.3$ and $J = 1.93 t$, $t_\perp = 1.42t$ (a);  $J = 1.6 t$, $t_\perp = 1.54t$ (b) and $J = 1.1 t$, $t_\perp = 1.78t$ (c). For Fig.~\ref{fig4SM}, we used $t_\perp=t=0.3$ and $J = 3t$, $J_\perp = 5t$ (a); $J = 2.67t$, $J_\perp = 4.67t$ (b) and $J = 2t$, $J_\perp = 4t$ (c). Studying the mixing of the ba-$s_\pm$-wave with a finite in-plane s-wave component, which is not shown, yields similar results as those in the multiorbital La-327 model.  Alternatively, a finite ratio $\Delta_\perp/\Delta_s$ in the $\Delta \rightarrow 0$ limit has to be allowed to vary freely in the full set of gap equations.

For the multiorbital La-327 model, we restricted the momentum summations to a 100 meV shell around the Fermi surface to reduce the numerical costs.  To check the consistency,  we applied this approximation to the single orbital model and verified that the suppression curves are minimally affected. We use $J = 2.36$ eV for the $d$-wave of the main text and $J_\perp = 1.93$ eV for the interlayer s-wave shown in Fig.~2(a) and Fig.~\ref{fig1SM}. For the extended s-wave in Fig.~\ref{fig1SM}, $J_\perp = 1.93$ eV and $J=(t/t_\perp)^2 J \approx 1.73 J$ are used. To shift the $\gamma$ pocket 50 meV below the Fermi surface, we change the orbital on-site energy for the d$z^2$-orbital from 0.459 eV to -0.095 eV. To achieve a fixed filling of 1.5, a chemical potential of $\mu\approx-0.307$ eV is necessary. For the interlayer ba-$s$-wave and d-wave shown in Fig.~2(b) and Fig.~\ref{fig3SM}, $J_\perp = 6.4$ eV and $J = 2.4$ eV are used, respectively. For the intralayer dominated s-wave in Fig.~\ref{fig3SM}, $J \approx 1.57$ eV and correspondingly $J_\perp=2.71$ eV are used.

\bibliography{literatur}